\documentclass[preprint, 3p]{elsarticle}
\usepackage{graphicx}
\graphicspath{{./}}

\usepackage{amssymb}
\usepackage[cmex10]{amsmath}
\usepackage{algorithmic}
\usepackage{array}

\usepackage{mdwmath}
\usepackage{mdwtab}
\usepackage{eqparbox}
\usepackage{multirow} 
\usepackage{fixltx2e}
\usepackage{stfloats}
\usepackage{url}

\usepackage[numbers]{natbib}

\usepackage{listings}
\usepackage{xcolor}
\usepackage{caption}
\usepackage{mathtools}
\journal{Computers and Electrical Engineering}
\begin{document}
\begin{frontmatter}

  \title{Efficient Resource Sharing Through GPU Virtualization on Accelerated High Performance Computing Systems\tnoteref{tnote}}
\tnotetext[tnote]{A preliminary version of this paper has been presented at 2011 IEEE International Conference on Parallel Processing (ICPP'11) \cite{c_vgpu}.} 
\author[gwu]{Teng Li\corref{cor1}}
\ead{tengli@gwu.edu}
\author[gwu]{Vikram K. Narayana}
\ead{vikramkn@ieee.org}
\author[gwu]{Tarek El-Ghazawi}
\ead{tarek@gwu.edu}
\cortext[cor1]{Corresponding author}   
\address[gwu]{NSF Center for High-Performance Reconfigurable Computing (CHREC), Deparment of Electrical and Computer Engineering, The George Washington University, 801 22nd Street NW, Washington, DC, 20052, USA}

\begin{abstract}
The High Performance Computing (HPC) field is witnessing a widespread adoption of Graphics Processing Units (GPUs) as co-processors for conventional homogeneous clusters. The adoption of prevalent Single-Program Multiple-Data (SPMD) programming paradigm for GPU-based parallel processing brings in the challenge of resource underutilization, with the asymmetrical processor/co-processor distribution. In other words, under SPMD, balanced CPU/GPU distribution is required to ensure full resource utilization. In this paper, we propose a GPU resource virtualization approach to allow underutilized microprocessors to efficiently share the GPUs. We propose an efficient GPU sharing scenario achieved through GPU virtualization and analyze the performance potentials through execution models. We further present the implementation details of the virtualization infrastructure, followed by the experimental analyses. The results demonstrate considerable performance gains with GPU virtualization. Furthermore, the proposed solution enables full utilization of asymmetrical resources, through efficient GPU sharing among microprocessors, while incurring low overhead due to the added virtualization layer.

\end{abstract}

\end{frontmatter}

\section{Introduction}
\label{intro}

Recent years have seen the proliferation of Graphics Processing Units (GPUs) as application accelerators in High Performance Computing (HPC) Systems, due to the rapid advancements in graphic processing technology over the past few years and the introduction of programmable processors in Graphics Processing Units (GPUs), which is also known as GPGPU, or General-Purpose Computation on Graphic Processing Units \cite{wp_gpgpu}. As a result, a wide range of HPC systems have incorporated GPUs to accelerate applications by utilizing the unprecedented floating point performance and massively parallel processor architectures of modern GPUs, which can achieve unparalleled floating point performance in terms of FLOPS (FLoating-point Operations Per Second) up to the TeraFLOP barrier \cite{wp_tesla} \cite{wp_amd13}. Such systems range from clusters of compute nodes to parallel supercomputers. While examples of GPU-based computer clusters can be found in academia for research purpose, such as \cite{fan2004gpu} and \cite{kindratenko2009gpu}. The contemporary offerings from supercomputer vendors have begun to incorporate professional GPU computing cards into the compute blades of their parallel computer products; example include the latest Cray XK7 \cite{m_crayxk7} and SGI Altix UV \cite{m_sgi} supercomputers. Yet another notable example is the Titan supercomputer \cite{w_titan} currently ranking the 2\textsuperscript{nd} in the Top 500 supercomputer list \cite{w_top500}. Titan is equipped with 18,688 NVIDIA Tesla GPUs and is thereby able to achieve a sustained 17.59 PFLOPS LINPACK performance \cite{w_top500}.

\begin{table}[t]
  \caption{GPU-based Supercomputers in The Top 30 List}
  \label{tb:sclist}
  \centering
  \begin{tabular}{l|l|l|l}
    \hline
    Supercomputer (Ranking)&\# of CPU Cores &\# of GPUs&CPU/GPU Ratio\\
    \hline
    Titan (2\textsuperscript{nd}) & 299,008 & 18688 & 16\\
    \hline
    Tianhe-1A (10\textsuperscript{th}) & 102,400 & 7,168 & 14.3\\
    \hline
    Nabulae (16\textsuperscript{th})& 55,680 & 4,640 & 12\\
    \hline
    Tsubame2.0 (21\textsuperscript{st})& 17,984 & 4,258 & 4.2\\
    \hline
  \end{tabular}
\end{table}
Development of parallel applications for any of these GPU-based heterogeneous HPC systems requires the use of parallel programming techniques composed of both CPU and GPU parallel programs to fully utilize the computing resources. Among the different parallel programming approaches, the most commonly followed programming approach is the Single Program Multiple Data (SPMD) model \cite{darema2001spmd}. Under the SPMD model, multiple processes execute the same program on different CPU cores, simultaneously operating on different data sets in parallel. Techniques of messaging passing such as MPI \cite{gropp1999using} are often deployed to achieve the SPMD parallelism effectively with necessary required inter-processor communication. By allowing autonomous execution of processes at independent points of the same program, SPMD serves as a convenient yet powerful approach for efficiently making use of the available hardware parallelism.

With the introduction of hardware accelerators, such as GPUs, as co-processors, HPC systems are exhibiting architectural heterogeneity that has given rise to programming challenges not previously existing in traditional homogeneous platforms. With the SPMD approach used for programming most of the homogeneous parallel architectures, directly offloading the program instances on to GPUs is not feasible due to the different Instruction Set Architectures (ISAs) of CPUs and GPUs. Moreover, GPUs are primarily suited for the Compute-Intensive portions of the program, serving as co-processors to the CPUs in order to accelerate these sections of the parallel program. The ``single program'' requirement of SPMD therefore means that every program instance running on the CPUs must necessarily have access to a GPU accelerator. In other words, it is necessary to maintain a one-to-one correspondence between CPUs and GPUs, that is, the number of CPU cores must equal the number of GPUs. However, due to the proliferation of many-core microprocessors in HPC systems, the number of CPU cores generally exceeds the number of GPUs, which is true for all GPU-based supercomputers in the top 20 list, as shown in Table \ref{tb:sclist} \cite{w_top500} \cite{w_titan} \cite{w_nscc} \cite{w_sugon} \cite{m_tsubame}. Therefore, the problem of system computing resources underutilization with the SPMD approach is general across GPU-based heterogeneous platforms and urgently needs to be solved with the increasing number of CPU cores in a single CPU due to the fast advancement of multi/many core technologies.

However, even though the number of physical GPUs can not match the number of CPU cores found in contemporary HPC systems, modern high-end GPU architecture is designed as massively parallel and composed of up to thousands of processing cores \cite{wp_amd}. Moreover, since GPU programs are composed of parallel threads executed in parallel on these many processing cores physically, it is possible to achieve execution concurrency of multiple GPU programs on the single GPU. The Fermi \cite{wp_fermi} architecture from NVIDIA consisting up 512 Streaming Processor (SP) cores allows concurrent execution of up to 16 GPU kernels \cite{m_cuda}. The increasing parallel computation capabilities of modern GPUs enables the possibility of sharing a single GPU to compute different applications or multiple instances of the same application, especially when the application problem size and parallelism is significantly smaller than the inherent parallelism capacity of the GPU.

In this paper, targeting at the problem of resource underutilization under SPMD model, we propose the concept of efficiently sharing the GPU resources among the microprocessor cores in heterogeneous HPC systems, by providing a virtualized unity ratio of GPUs and microprocessors. We develop the GPU resource virtualization infrastructure as the required solution for asymmetrical resource distribution. The GPU virtualization infrastructure provides the required symmetry between GPU and CPU resources by virtually increasing the number of GPU resources, thereby enabling efficient SPMD execution. In the meantime, we also theoretically analyze the GPU program execution within a single compute node, and provide analytical execution models for our resource virtualization scenarios, which provide different extents of GPU kernel concurrency support. Furthermore, we conduct a series of experiments using our virtualization infrastructure on our NVIDIA GPU cluster node as verifications of the proposed modeling analysis as well as demonstration of well-improved process-level GPU sharing efficiency using the proposed virtualization approach.

The rest of this paper is organized as follows. Section \ref{rw} provides an overview of related work on GPU resource sharing in heterogeneous HPC systems and GPU virtualization under different contexts. A background of current GPU architectural and programming model is given in Section \ref{bg}, followed by a formal analysis of the GPU execution models for our proposed virtualization solution in Section \ref{model}. Section \ref{infra} then discusses the implementation details of the GPU resource sharing and virtualization infrastructure. The experimental results are presented and discussed in Section \ref{exp}, followed by our conclusions in Section \ref{conc}.

\section{Related Work}
\label{rw}

With the continued proliferation of GPGPU in the HPC field, virtualization of GPUs as computing devices is gaining considerable attention in the research community. One stream of recent research has focused on providing access to GPU accelerators within virtual machines (VMs). \cite{gupta2009gvim} \cite{shi2009vcuda} \cite{giunta2010gpgpu} Research studies in this stream focus on providing native CUDA programmability support within VMs, on computing systems equipped with NVIDIA GPUs. Gupta et al. \cite{gupta2009gvim} presented their GViM software architecture, which allows CUDA applications to execute within VMs under the Xen virtual machine monitor. The GPU is controlled by the management OS, called ``dom0'' in Xen parlance. The actual GPU device driver therefore runs in dom0, while applications execute on the guest OS or VM. By using a split-driver approach, the CUDA run-time API function calls from the application are captured by an interposer library within the VM, followed by transfer of the function parameters to dom0 for actual interaction with the GPU. A similar approach is adopted in vCUDA by Shi et al. \cite{shi2009vcuda}, also on Xen. They additionally provide suspend and resume facility by maintaining a record of the GPU state within the virtualization infrastructure in the guest OS as well as the management OS domain.

However, the primary drawback of providing GPU access by using Xen is that NVIDIA drivers do not officially support Xen and CUDA drivers with version later than 2.2 do not work under Xen, which therefore makes the aforementioned approach unportable. Giunta et al. \cite{giunta2010gpgpu} circumvent this problem by using the Kernel Virtual Machine (KVM) available in Linux distribution, while following a similar split-driver approach coupled with API call interception. Their software infrastructure, termed gVirtuS, focuses on providing GPU access to VMs within virtual clusters, as part of a cloud computing environment. For cases when the host machine does not have a local GPU, they envision that the virtual machines will use TCP/IP to communicate with other hosts within the virtual cluster in order to gain access to remote GPUs.  

While achieving GPU virtualization by providing CUDA support within VMs appears to be an attractive solution for our problem, launching multiple VMs within the same compute node can result in significant overheads for HPC applications targeting at performances. To elaborate, efficient SPMD execution requires all CPU cores (processes) to have a virtual view of a GPU; with the VM approach, this would require a VM to be launched for every CPU core within the compute node. Since the number of CPU cores per node is rapidly on the rise, significantly increasing overheads with the VM-based approach for GPU sharing are inevitable. Further, all the available VM-based GPU sharing approaches time-share the GPU. Thus, the potential for simultaneous executing GPU functions from multiple processes can not been exploited.

Other types of solutions have also been proposed for GPU virtualization in HPC systems. For example, Duato et al \cite{duato2010efficient} propose the use of remote GPU access similar to \cite{giunta2010gpgpu}, for cases when high performance clusters do not have a GPU within every compute node. Instead of using a VM, they propose a GPU middleware solution consisting of a daemon running on GPU-capable nodes that serves requests from non-GPU node clients. The client nodes incorporate a CUDA wrapper library to capture and transfer API function calls to the server using TCP/IP. Although the VM overheads are removed, their proposed solution can result in communication overheads in accessing GPUs from remote compute nodes. Moreover, simultaneous execution of multiple GPU kernels is not discussed.

Another stream of research has concentrated on GPU sharing to eliminate resource under-utilization. Guevara et al. \cite{guevara2009enabling} propose an approach that involves run-time interception of GPU invocations from one or more processes, followed by merging them into one GPU execution context. Currently their solution is demonstrated for two kernels, with the merged kernels predefined manually. Similarly, Saba et al. \cite{saba2010anytime} presented an algorithm that
allocates GPU resources for tasks based on the resource goals and workload size. While they target at a different problem of a time bound algorithm that optimizes the execution path and output quality, the employed GPU sharing approach is to merge the kernels similarly. Although kernel merging can be useful for SPMD execution, it would need compiler support for generation of the combined kernels a priori, which can be avoided by using the concurrent kernel execution support from Fermi. Moreover, with merged kernels, multiple kernels are ¿invoked¿ simultaneously, therefore it does not exploit the possibility to hide data transfer overhead with kernel execution, which can be achieved through concurrent kernel execution and data transfer from current GPU architecture. Similar in scheduling purpose as \cite{guevara2009enabling}, our previous work \cite{c_gpusch, cf14, icpads15} described an off-line scheduling framework to achieve improved performance and resource utilization of many GPU tasks with possible inter-task concurrencies through GPU sharing. While \cite{c_gpusch} targets at a completely different problem, it utilizes the similar analytical GPU sharing approach as our proposed work in this paper. Meanwhile, GPU resource under-utilization has also been studied in cloud computing. Ravi et al. \cite{ravi2011supporting} proposed a GPU sharing framework for the cloud environment, base on which they provided a GPU kernel consolidation algorithm to combine kernel workloads within the cloud and thus achieves improved throughput by utilizing both space and time sharing.

For thread-level GPU sharing, Peters et al. \cite{peters2010efficiently} proposed another technique for sharing an NVIDIA GPU among multiple host threads on a compute node. Their technique involves the use of persistent kernels that are initialized together within a single execution context, and remain executing on the GPU indefinitely. Each of these persistent kernels occupies a single thread block on the GPU, and they execute the appropriate function based on commands written to the GPU memory by the host process running on the CPU. This ``management thread'' accepts requests from other CPU threads for the use of GPU resources. Their approach allows for multiple kernels to simultaneously execute even on devices that do not support concurrent kernel execution. However, due to the persistent nature of the kernels, the number of thread blocks is severely limited, which means that the memory latency may not be hidden effectively. As a result, highly data-parallel applications will not be able to take full benefit of the GPU resources. Furthermore, their approach requires significant changes to the application code to fit within a thread block. Also, the communication mechanism between the management thread and other threads is not clear. Nevertheless, the use of a management process to control the GPU and manage the resources is a common feature shared with our proposed solution.

To provide process-level GPU sharing, our initial work \cite{c_vgpu} presented the GPU virtualization approach and prototype as the basis for our proposed work in this paper. Another solution is proposed by the S\_GPU project \cite{wp_sgpu}, which provides a software layer that resides between the application and the GPU, typically used for GPU time-sharing between MPI processes in a parallel program. Each MPI process is provided the view of a private GPU, through a custom stream-based API. Each process inserts the GPU commands, such as memory copy, kernel launch, etc, in the required sequence into a stream object, irrespective of the number of GPUs available. When the process initiates the execution of a stream, all the enqueued GPU commands are then executed in the required sequence. The S\_GPU software stack takes care of sharing the available GPUs among the streams from multiple processes. The approach followed by S\_GPU is complementary to \cite{c_vgpu} and our approach here, and may be combined with our proposed approach by simultaneously executing kernels from multiple processes for efficient GPU sharing. Furthermore, the efficient GPU sharing approach proposed by \cite{cf12, computers, dis} is relavant to this work as well.

On the other hand, virtualization of co-processors has been studied for other technologies as well, such as Field-Programmable Gate Arrays (FPGAs). For example, Huang and Hsiung \cite{huang2009hardware} provide an OS-based infrastructure that allows multiple hardware functions from different processes to be configured in the FPGA, by using the partial run-time reconfiguration feature. Their virtual hardware mechanism allows a given hardware function to be used by multiple applications, as well as enabling one application to access multiple hardware functions simultaneously in the FPGA. Similar FPGA work also includes the RAD architecture proposed by \cite{rad}. Our previous work \cite{el2008virtualizing} also uses the partial run-time reconfiguration feature of the FPGA, albeit for enabling efficient SPMD execution in HPC systems. This work partitions the FPGA into multiple regions, and allocates each region to a CPU core within the compute node. By thus providing a virtual FPGA to every CPU core, the required 1:1 ratio of the number of CPU cores to the number of virtual FPGAs is achieved. This work forms the basis of our proposed virtualization solution for GPU-based HPC systems. Our proposed approach effectively utilizes modern GPU features, such as concurrent kernel execution and concurrent data transfer and execution, to improve the execution performance of the system. Overheads are kept small by maintaining a simple communication mechanism between the CPU processes and our proposed virtualization infrastructure.

\begin{figure}[t]
  \centering  
    \includegraphics[width=0.8\linewidth]{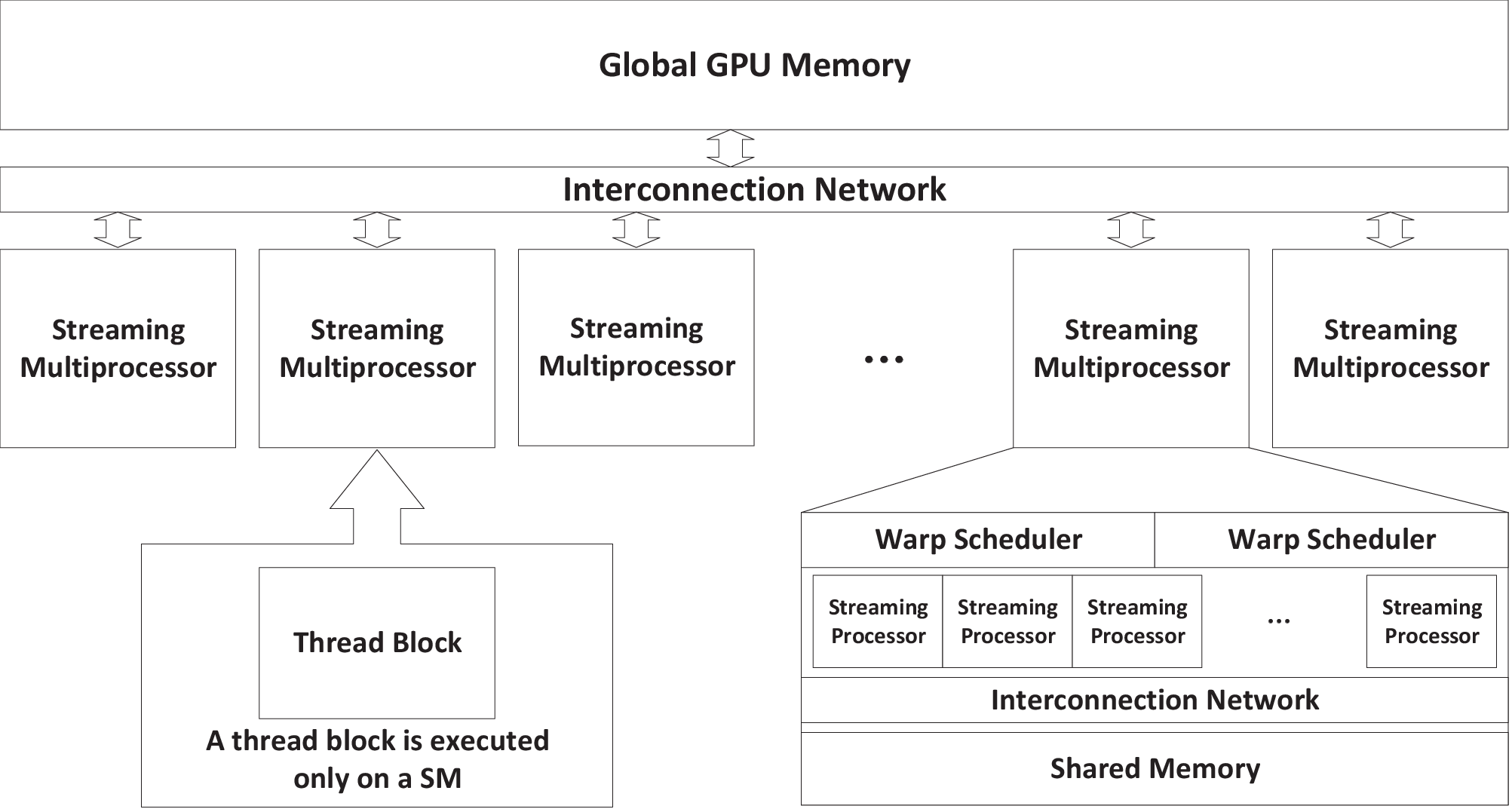}
    \caption{An Overview of Current GPU Architecture}
    \label{fig:gpu_background}
\end{figure}

\section{Background on GPU Programming and Architectural Model}
\label{bg}

A brief overview of GPU programming and computing architectural model is presented in this section. Our further modeling analysis and implementation are based on Compute Unified Device Architecture (CUDA) \cite{m_cuda} and Open Computing Language (OpenCL) \cite{m_opencl} on the programming model side as well as the NVIDIA GPU architecture \cite{wp_fermi} on the hardware architecture side.

\subsection{Programming Models of Modern GPUs}
As previously mentioned, CUDA and OpenCL are two prevalent GPU computing programming models provided by NVIDIA and Khronos Working Group, respectively. Both models follow Single Instruction Multiple Thread (SIMT) model by executing data-parallel GPU kernel functions within the GPU, while providing abstractions of thread group hierarchy and shared memory hierarchy. In terms of thread group hierarchy, both models (CUDA/OpenCL) provide three hierarchy levels: Grid/NDRange, Block/Workgroup and Thread/Workitem. For convenience, we will use CUDA terminology in the rest of this paper, namely, grids, blocks and threads. GPU kernels are launched per grid and a grid is composed of a number of blocks, which have access to the global device memory. Each block consists of a group of threads, which are executed concurrently and share the accesses to the on-chip shared memory. Each thread is a very lightweight execution of the kernel function. From programming perspective, the programmer needs to write the kernel program for one thread and decides the total number of threads to be executed on the GPU device while dividing the threads into blocks based on the data-sharing pattern, memory sharing and architectural considerations.

\subsection{Architectural Model of Modern GPUs}

Modern GPUs are composed of massively parallel processing units and hierarchies of memories. From the architectural perspective of modern GPU hardware, the GPU architecture (Fermi \cite{wp_fermi}) from NVIDIA, as shown in Figure \ref{fig:gpu_background}, is composed of up to 16 Streaming Multiprocessors (SMs), each of which can be further decomposed of 32 Streaming Processor (SP) cores as the second processing element hierarchy. Memory hierarchies are composed of the global device memory shared by all SMs and private shared memories of each SM that are shared by all SPs within each SM. Each thread is executed on an SP core and each block runs on an SM at a given time. While each SP holds a number of registers for each thread and executes threads sequentially; within each SM, threads are scheduled as a batch of 32 threads called a warp. Two warp schedulers exist in an SM and multiple warps are allowed to co-exist on the same SM. Especially when threads have high memory access latency, increasing the warp occupancy by having multiple warps co-exist on a single SM simultaneously can improve the overall execution performance.

\subsection{Kernel-level Execution Flow of Modern GPUs}

Upon the launching of kernel composed of multiple blocks, each block can only be executed on an SM. Multiple blocks can only reside on the same SM provided there are enough SM resource, which includes the register, shared memory and total warps can be scheduled in an SM . Since each SM has limited number of registers, a fixed size of shared memory and a maximum number of warps that can co-exist and be scheduled, only multiple blocks meeting these constraints can be scheduled within a single SM. However, blocks are only limited to a single kernel. In other words, blocks from different kernels can not be concurrently scheduled in conventional GPU devices. Nevertheless, current CUDA devices with computing capability higher than 2.0 (Fermi or later) support concurrent kernel execution, which allows different
kernel to be launched from the same process (GPU program) using CUDA streams. Concurrent kernel execution allows blocks from different kernels to be scheduled simultaneously. Furthermore, by using asynchronous CUDA streams, concurrent data transfer and kernel execution can also be achieved among multiple streams, each of which carries a GPU kernel. Note, however, that GPU kernels launched from independent CPU processes cannot be concurrently executed, which necessitates the proposed virtualization approach and implementation described in the following sections.

\begin{figure}[t]
  \centering  
    \includegraphics[width=0.7\linewidth]{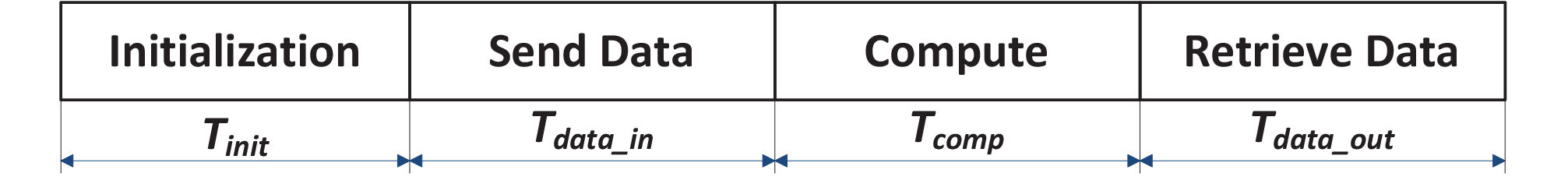}
    \caption{The GPU Execution Cycle for a Single Process}
    \label{fig:stages}
\end{figure}
\begin{figure}[t]
  \centering  
    \includegraphics[width=1\linewidth]{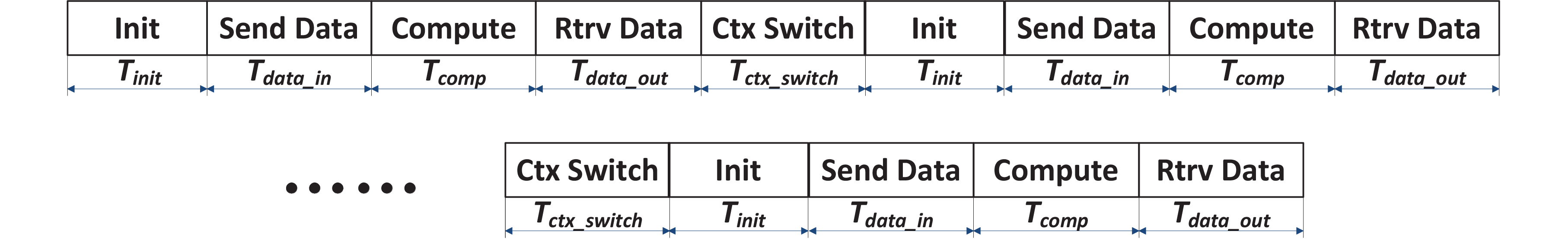}
    \caption{The Native GPU Execution Model from Multiple Processes without Virtualization}
    \label{fig:seq}
\end{figure}
\section{Process-level Sharing Modeling and Performance Analysis with GPU Virtualization}
\label{model}

In this section, we perform a theoretical study on the performance of process-level GPU sharing. We describe and analyze the achievable GPU sharing efficiency when our virtualization approach is applied. Our analysis is based on the actual GPU sharing scheme provided by GPU virtualization. Here we first perform a study of the factors that affect the overall performance of multiple GPU kernel executions. By conducting this study, we will be able to investigate the performance potential of using our proposed concept of centralized daemon-process-based GPU sharing. We will first layout the conventional scheme when multiple processes share a single GPU, followed by potential improvement technique using GPU virtualization. Our analysis is primarily based on a series of GPU kernel execution models depicting different process-level kernel execution scenarios. The proposed execution model considers multiple factors affecting the overall sharing efficiency and captures the timing of multiple kernel executions from many GPU requests on the process level (from many processors). Furthermore, we define necessary model parameters and derive the corresponding performance formula with performance comparison and analysis.

To study the potential GPU performance gain with the virtualization approach, we primarily focus our discussions on GPU tasks while excluding the CPU tasks. This is because we are mostly interested in enhancing the execution performance of many GPU tasks launched through SPMD programs. Meanwhile, we also try to avoid adding unnecessary complications to model the scenario of co-scheduling both CPU and GPU tasks.

\subsection{Conventional Process-level GPU Request and Execution Model}


Conventional, when multiple GPU kernel execution requests are launched from many SPMD processes, processes have direct access to the physical GPU without the proposed add-on virtualization layer. Here we base our analysis on a simplified scenario that a single GPU is requested from multiple processes running on multi-processors (cores) within each cluster node, due to the aforementioned asymmetrical CPU/GPU resource distribution. Under this scenario, multiple GPU requests from processors within each node will place individual GPU tasks/kernels on the shared single physical GPU simultaneously. However, current GPU device is only able to accept process-level requests one after another and thus sequentially execute multiple GPU kernels.

To analyze the execution pattern of conventional GPU sharing, we model the execution cycle for a given process to perform a GPU computing task in Figure \ref{fig:stages}. The modeled execution cycle is composed of the following four timing stages: the process first initializes the GPU device, creates its own GPU context and allocates the GPU device memory; followed by sending the input data, computing the task and retrieving the data back to the request process. Figure \ref{fig:stages} also specifies the timing parameters for each stage of the execution cycle. Table \ref{tb:para} further defines necessary parameters used for the following modeling analysis.

\begin{table}[t]
  \caption{Parameters Defined for Analytical Modeling}
  \label{tb:para}
  \centering
  \begin{tabular}{@{}l@{}|@{}p{14.5cm}@{}}
	\hline
	\textbf{Symbol} & \textbf{Definition} \\
    	\hline
	\emph{N\textsubscript{processor}} & The total number of processors in each node\\
	\hline
	\emph{N\textsubscript{process}} & The number of parallel SPMD processes (GPU tasks) running in each node, which should not exceed \emph{N\textsubscript{processor}}\\
	\hline
	\emph{N\textsubscript{VGPU}} & The number of virtual GPU resources for SPMD processes exposed from the virtualization layer, which equals to \emph{N\textsubscript{processor}}\\
	\hline
	\emph{T\textsubscript{init}} & The time for each process to initialize the GPU and corresponding resources\\
	\hline
	\emph{T\textsubscript{ctx\_switch}} & The average GPU context-switch overhead associated with each process\\
	\hline
	\emph{T\textsubscript{data\_in}} & The time for each process to transfer the input data into the GPU device memory\\ 
	\hline	
	\emph{T\textsubscript{data\_out}} & The time for each process to retrieve the result data back from the device memory\\
	\hline
	\emph{T\textsubscript{comp}} & The time for the GPU to compute the task\\
	\hline
	\emph{T\textsubscript{total\_ci\_ps1}} & The time to execute C-I kernels from all processes under PS-1 with virtualization\\
	\hline
	\emph{T\textsubscript{total\_ci\_ps2}} & The time to execute C-I kernels from all processes under PS-2 with virtualization\\
	\hline
	\emph{T\textsubscript{total\_ioi\_ps1}} & The time to execute IO-I kernels from all processes under PS-1 with virtualization\\
	\hline
	\emph{T\textsubscript{total\_ioi\_ps2}} &The time to execute IO-I kernels from all processes under PS-2 with virtualization\\
	\hline
	\emph{T\textsubscript{total\_no\_vt}} & The time to execute kernels from all processes without virtualization\\ 
	\hline
	\emph{S\textsubscript{ci}} & The theoretical speedup that can be achieved for C-I kernels\\
	\hline
	\emph{S\textsubscript{ioi}} & The theoretical speedup that can be achieved for IO-I kernels\\
	\hline
	\emph{S\textsubscript{max\_ci}} & The theoretical maximum speedup that can be achieved for C-I kernels\\
	\hline
	\emph{S\textsubscript{max\_ioi}} & The theoretical maximum speedup that can be achieved for IO-I kernels\\
	\hline
  \end{tabular}
\end{table}

\begin{figure*}[t]
\begin{lstlisting}[caption=CUDA Stream Programming Style-1 (PS-1), label=ps1]
for (int i = 0; i < NumStream; i++)
    cudaMemcpyAsync (dst, src, count, cudaMemcpyHostToDevice, stream[i]);
for (int i = 0; i < NumStream; i++)
    myKernel <<<GridSize, BlockSize, ShareMemSize, stream[i]>>> (parameter);
for (int i = 0; i < NumStream; i++)
    cudaMemcpyAsync (dst, src, count, cudaMemcpyDeviceToHost, stream[i]);
\end{lstlisting}
\end{figure*}

\begin{figure*}[t]
\begin{lstlisting}[caption=CUDA Stream Programming Style-2 (PS-2), label=ps2]
for (int i= 0; i < NumStream; i++) 
{
    cudaMemcpyAsync (dst, src, count, cudaMemcpyHostToDevice, stream[i]);
    myKernel <<<GridSize, BlockSize, ShareMemSize, stream[i]>>> (parameter);
    cudaMemcpyAsync (dst, src, count, cudaMemcpyDeviceToHost, stream[i]);
}
\end{lstlisting}
\end{figure*}

\begin{figure*}[t]
  \begin{minipage}[b]{0.5\linewidth}  
  \centering  
    \includegraphics[width=1\linewidth]{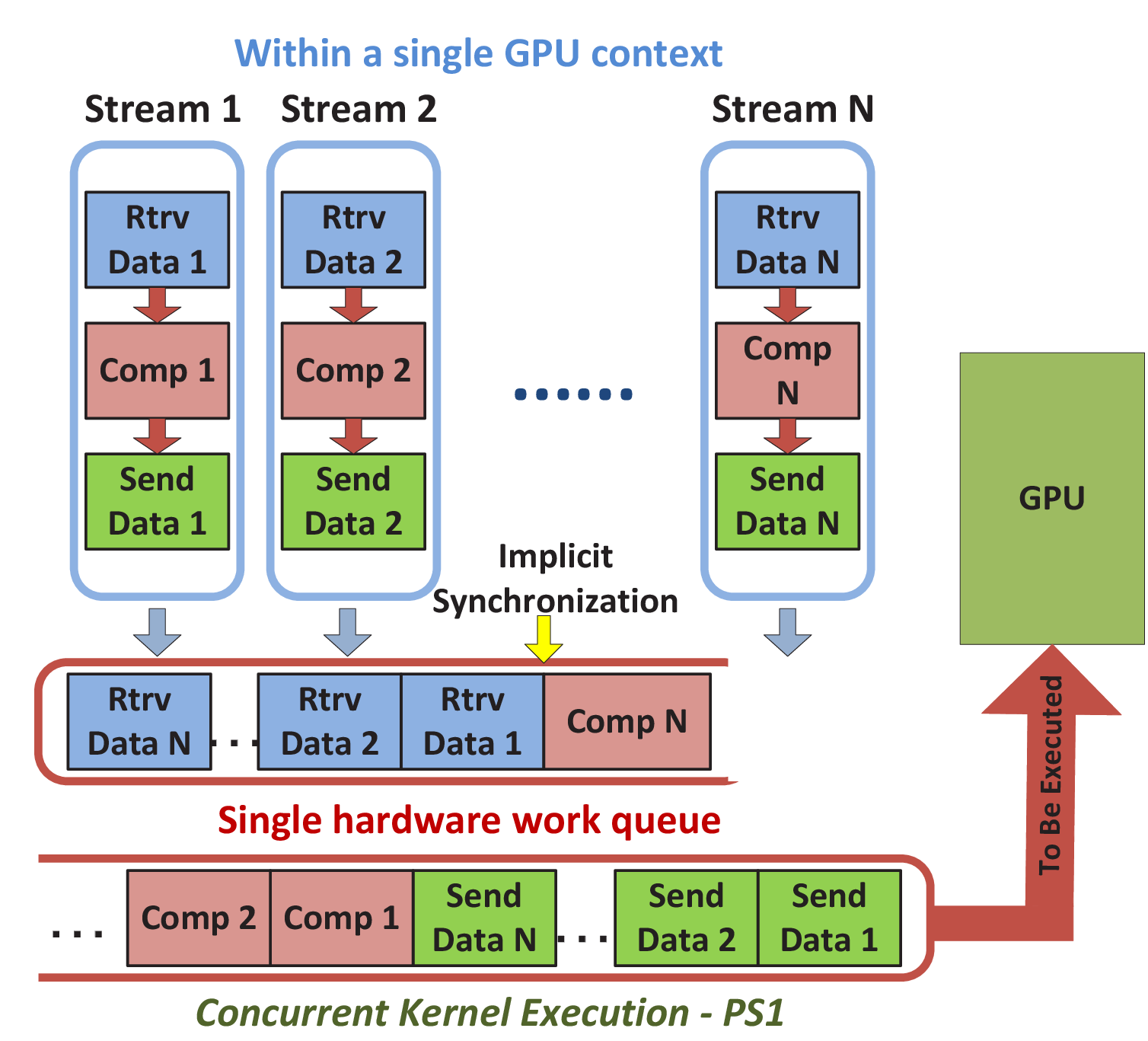}
    \caption{The Streaming Model for Kernel Concurrency}
    \label{fig:q-ps1}
\end{minipage}
\begin{minipage}[b]{0.5\linewidth}
  \centering  
    \includegraphics[width=1\linewidth]{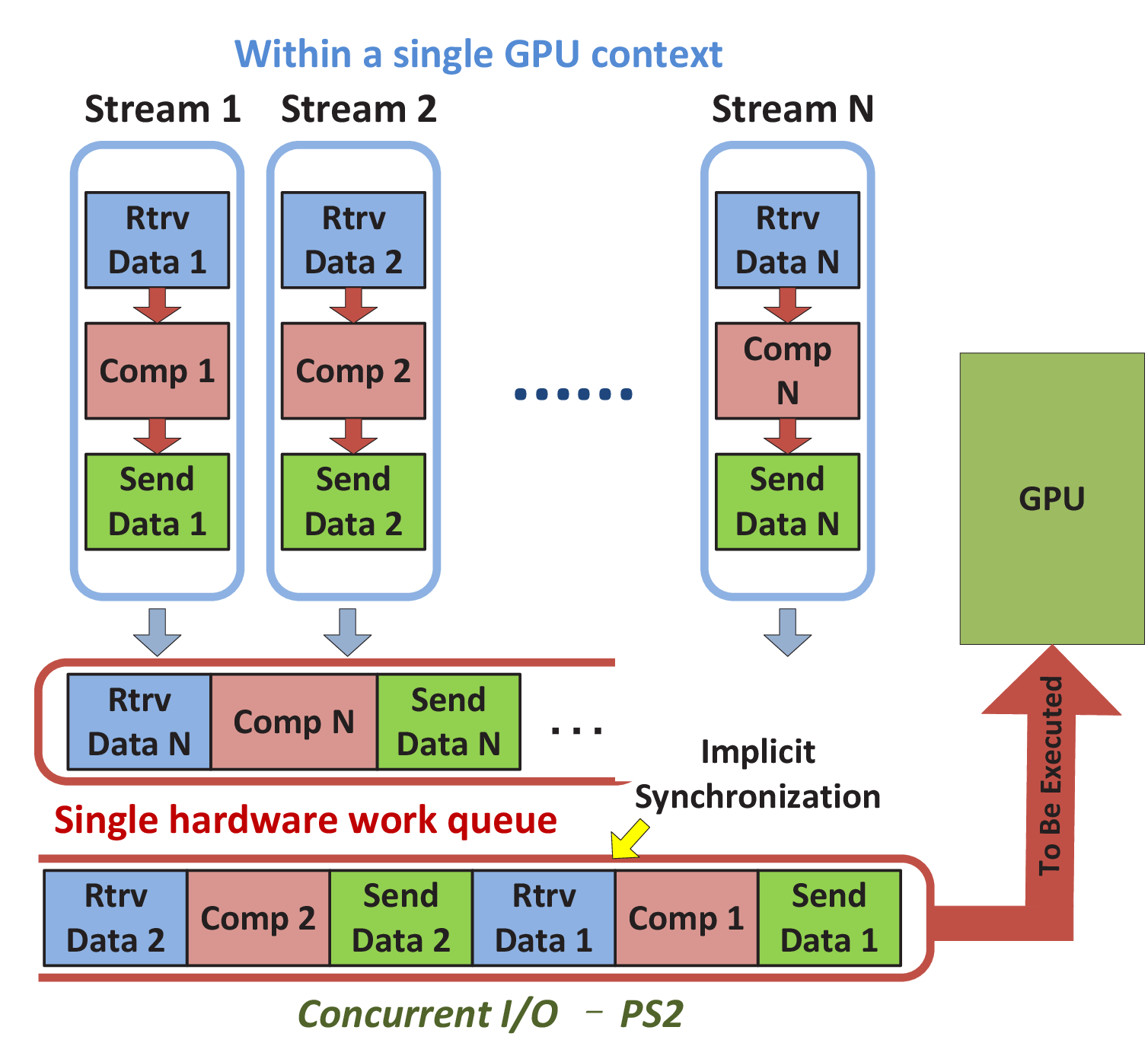}
    \caption{The Streaming Model for I/O Concurrency}
    \label{fig:q-ps2}
\end{minipage}
\end{figure*}
Under conventional execution scheme within each node, processes treat the GPU as a non-virtualized resource. Current NVIDIA CUDA architecture allows multiple host processes to be executed native sharing mode, by creating multiple GPU contexts for each process. The GPU kernels from multiple processes are serially executed in the order queued with respective GPU context-switch overheads. In other words, the execution model for conventional GPU sharing (with CUDA) among multiple processes is sequential among tasks with no concurrency possible. Moreover, each process needs an additional amount of time overhead to initialize the GPU device and its own resource such as context. Thus, in the execution model, we assume there is an average fixed GPU initialization overhead for each process, \emph{T\textsubscript{init}}, at the beginning. We also make assumption that there is an average context-switch overhead \emph{T\textsubscript{ctx\_switch}} between every two processes, as shown in Figure \ref{fig:seq}. Therefore, the modeled total execution time for the conventional scheme without virtualization can be derived with Equation \eqref{eq:seq}, which will be our performance comparison/analysis basis.
\begin{align}
  \label{eq:seq}
  \emph{T\textsubscript{total\_no\_vt}}=\emph{N\textsubscript{process}}(\emph{T\textsubscript{init}} +\emph{T\textsubscript{data\_in}}+\emph{T\textsubscript{comp}}+\emph{T\textsubscript{data\_out}})+(\emph{N\textsubscript{process}}-1)\emph{T\textsubscript{ctx\_switch}}
\end{align}

\subsection{The Optimized Sharing Scheme and Execution Model Under GPU Virtualization}

\subsubsection{Streaming Kernel Execution with CUDA and Potential Concurrencies}
As discussed earlier, GPU kernel execution from multi-processes achieves zero concurrency, however, current GPU technology (CUDA) is able to provide concurrencies of both kernel execution and I/O transfer through CUDA streams. However, the prevalent GPU Fermi architecture requires all CUDA streams to be launched simultaneously within a single GPU context to achieve both kernel and I/O concurrencies. In other words, if multiple kernels from SPMD processes can be launched through multiple CUDA streams from a single GPU context, the GPU can be efficiently shared among multiple kernels through streaming execution.

With streaming execution, three types of concurrencies are possible: (a) the execution of multiple concurrent kernels, (b) the concurrency between kernel execution and either GPU input or output data transfer, and (c) the concurrency between input data transfer and output data transfer. Currently, CUDA streams can be programmed with two styles \cite{m_cuda} achieving different concurrency/overlapping behaviors (kernel concurrency vs. I/O concurrency). For simplicity, we here define that Programming Style-1 (PS-1) is primarily to achieve kernel execution concurrency while Programming Style-2 (PS-2) is to focus on I/O concurrency. Both definitions will be used in the following discussions. To demonstrate the difference, here we list two simple code snippets for both PS-1 and PS-2 in Listing \ref{ps1} and \ref{ps2} respectively. Under both programming styles, \emph{NumStream} streams are launched. While the three stages: ``Send Data'',``Compute'' and  ``Rtrv Data'' are executed as a batch for each single stream consecutively under PS-2, the same stages of all streams are executed as a batch under PS-1.

Under Fermi architecture, up to 16-way concurrency of kernel launches are permitted from individual streams. Upon being executed on the GPU, all CUDA streams are multiplexed into a single hardware work queue, as demonstrated in both Figure \ref{fig:q-ps1} and \ref{fig:q-ps2}. Thus the two programming styles result in two different execution sequences inside the hardware work queue and thus different concurrencies due to data dependencies. We note that the three stages within each stream are all asynchronous operations. In other words, each operation is non-blocking provided there is no required data dependency and implicit synchronization.

Due to the separated for loops under PS-1, shown in List \ref{ps1}, ``Send Data'' from all streams are first queued into the hardware work queue, followed by each ``Comp'' from all streams, with each ``Rtrv Data'' followed at last, as shown in Figure \ref{fig:q-ps1}. For PS-1, the big for loop shown in List \ref{ps2} allows each stream sequentially following each other within the hardware work queue to be processed by the GPU.

Since we here mainly focus on an SPMD program, there is no data dependency among each of ``\emph{Send Data i}'' (\emph{i} ranges from \emph{1} to \emph{N}), while this also applies among each of ``\emph{Comp i}'' as well as each of ``\emph{Rtrv Data i}''. However, for each stage of a single stream \emph{i}, ``\emph{Comp i}'' has the data dependency on the input data of ``\emph{Send Data i}'' while ``\emph{Rtrv Data i}'' depends on the result data from ``\emph{Comp i}''. Therefore, PS-1 achieves different concurrency behavior from PS-2.

CUDA currently only provides limited stream support \cite{m_cuda}. Specifically, operation requires an implicit synchronization (dependency check) to see if a kernel has finished (``\emph{Rtrv Data i}'' for example) can: (1) start only when all prior kernel launches have started executing. (2) block all later kernel launches until checking on current kernel launch is completed. Thus, in Figure \ref{fig:q-ps1} under PS-1, ``\emph{Rtrv Data 1}'' can only start after ``\emph{Comp N}'' and achieve very little concurrency with ``\emph{Comp N}'' while concurrencies can be achieved between ``\emph{Send Data i}'' and ``\emph{Comp i}'' as well as among all ``\emph{Comp i}''s. Similarly, in Figure \ref{fig:q-ps2} under PS-2, ``\emph{Rtrv Data i}'' blocks ``\emph{Comp i+1}'' due to the required dependency check. Therefore ``\emph{Comp i+1}'' can only start after ``\emph{Comp i}'' is finished while `\emph{Send Data i+1}'' can still overlap with ``\emph{Rtrv Data i}'' and even ``\emph{Comp i}''. Those overlapping behaviors serve as the basis for our execution model. 

\begin{figure}[t]
    \centering
    \includegraphics[width= 0.6\linewidth]{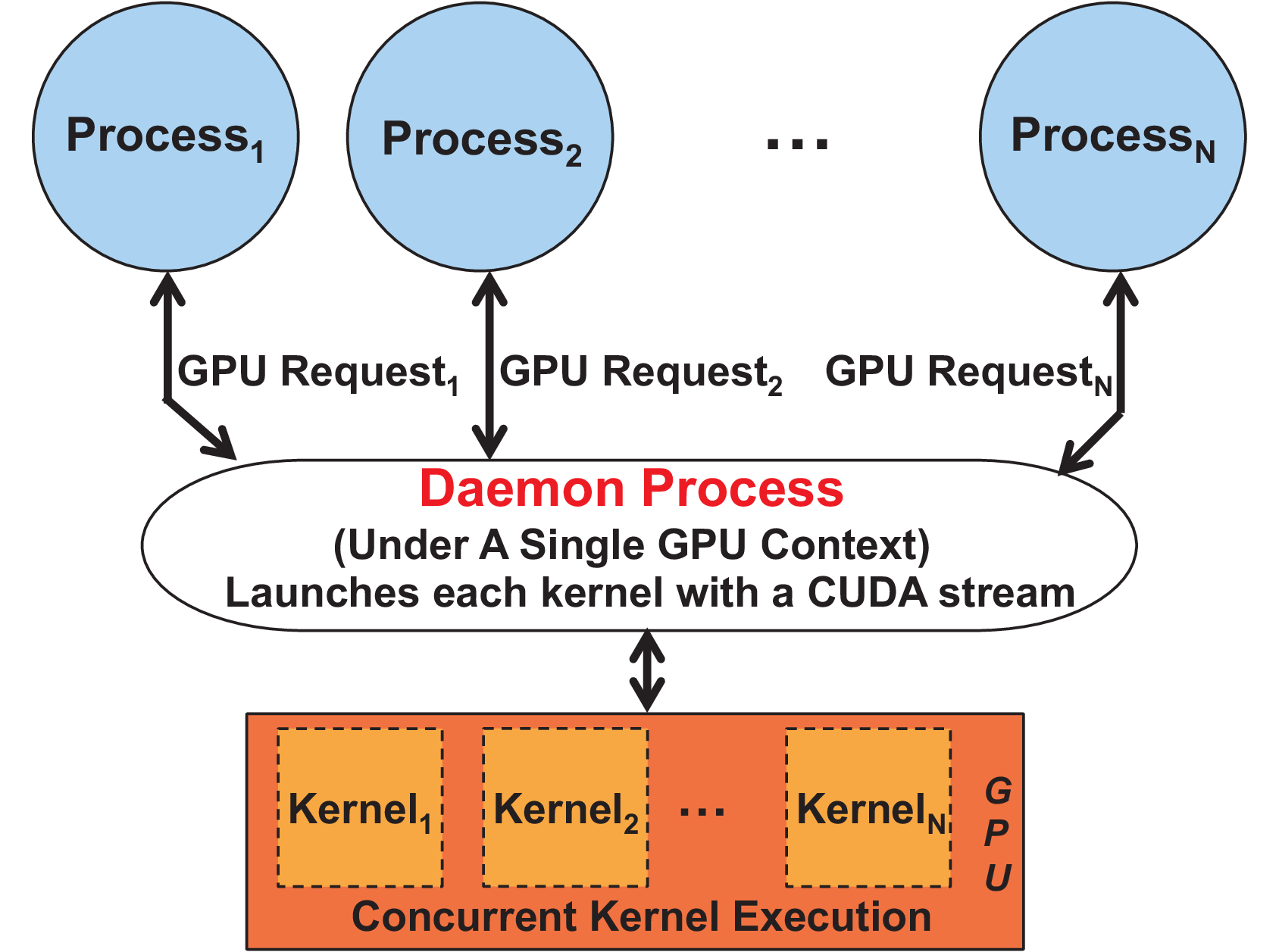}
    \caption{The Optimized Process-level GPU Sharing Scheme}
    \label{fig:dprocess}
 \end{figure}
\subsubsection{The Optimized Process-level GPU Sharing Scheme with GPU Virtualization}

Considering the possible performance advantage with execution concurrency, here we propose the process-level GPU sharing scheme utilizing streaming execution. Under the optimized GPU sharing scheme, requests from multiple SPMD processes are launched using separate CUDA streams within a single GPU context. Our virtualization technique creates a single daemon process that handles GPU requests from multiple processes, as shown in Figure \ref{fig:dprocess}. Since the daemon process creates a single required GPU context, launching GPU kernel requests inside the daemon process with separate CUDA streams allows the potential concurrencies among multiple processes. Additionally, the optimized sharing scheme is able to avoid context-switch overheads among processes as well as hide the GPU initialization overhead, which is due to the single daemon process been created. Further virtualization implementation details will be discussed in the following section.   
\begin{figure*}[t]
    \centering
    \includegraphics[width= 0.65\linewidth]{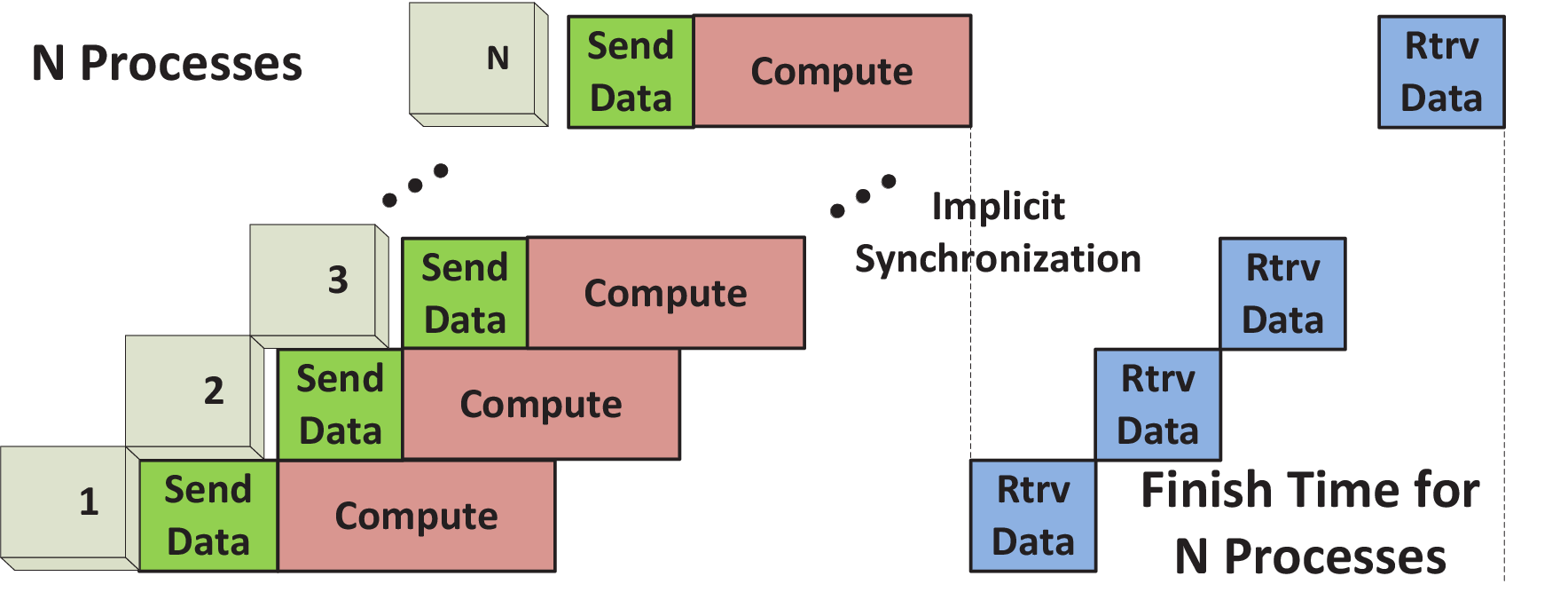}
    \caption{Compute-Intensive Kernels Under PS-1}
    \label{fig:ci-ps1}
 \end{figure*}
\begin{figure*}[t]
    \centering
    \includegraphics[width= 0.75\linewidth]{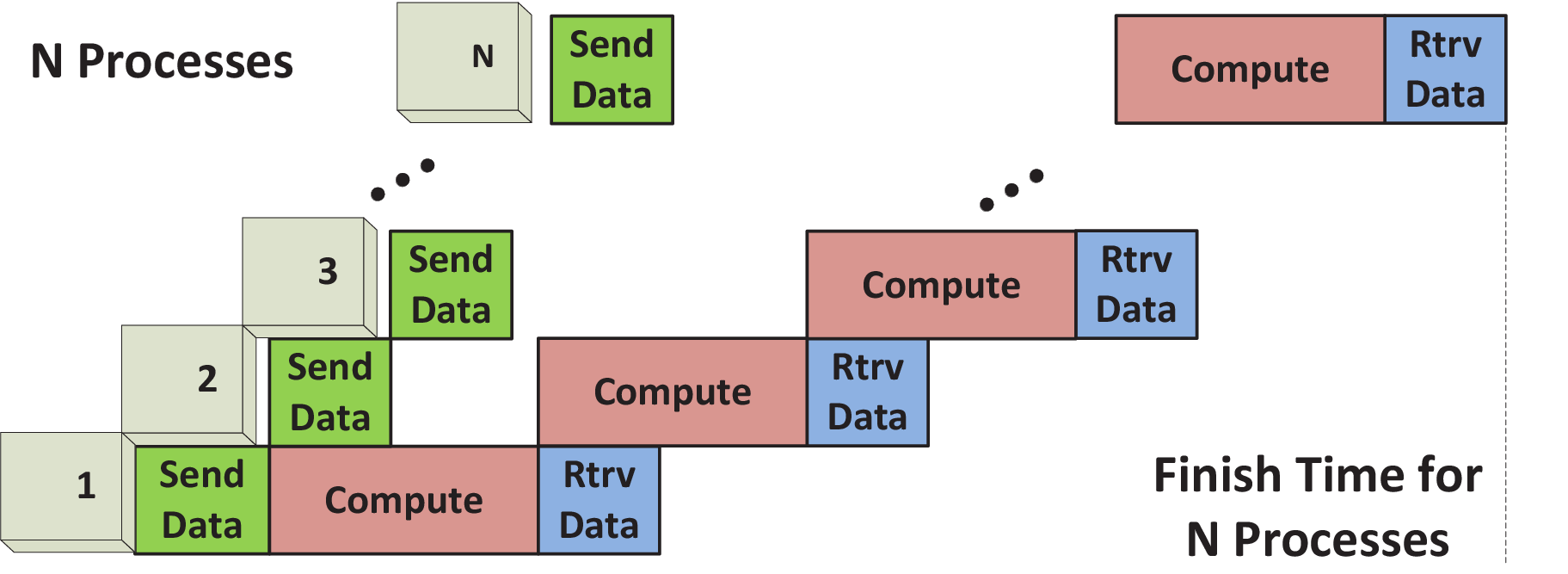}
    \caption{Compute-Intensive Kernels Under PS-2}
    \label{fig:ci-ps2}
 \end{figure*}
\begin{figure*}[t]
    \centering
    \includegraphics[width= 1\linewidth]{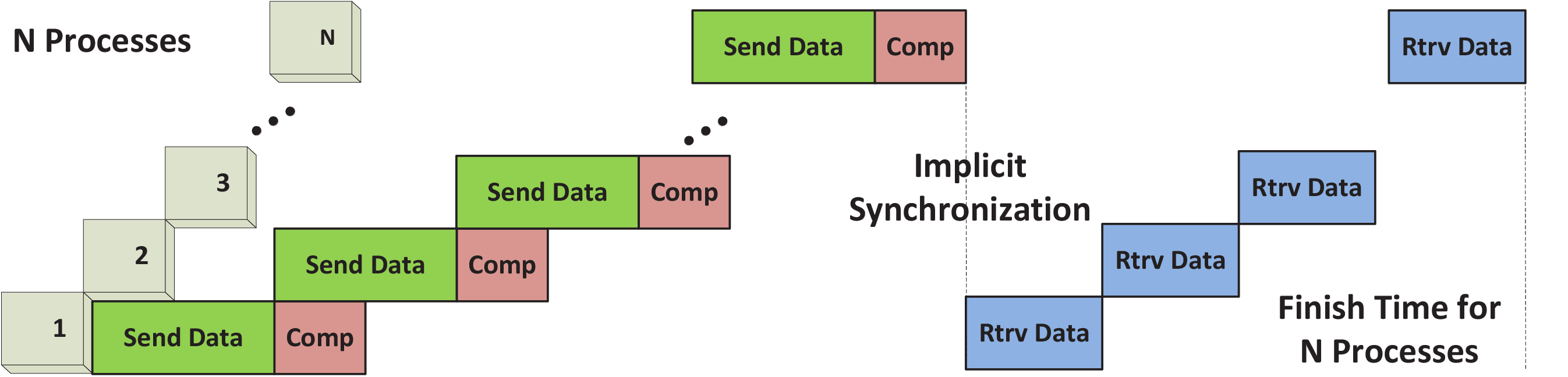}
    \caption{I/O-Intensive Kernels Under PS-1}
    \label{fig:ioi-ps1}
 \end{figure*}
\begin{figure*}[t]
    \centering
    \includegraphics[width= 0.8\linewidth]{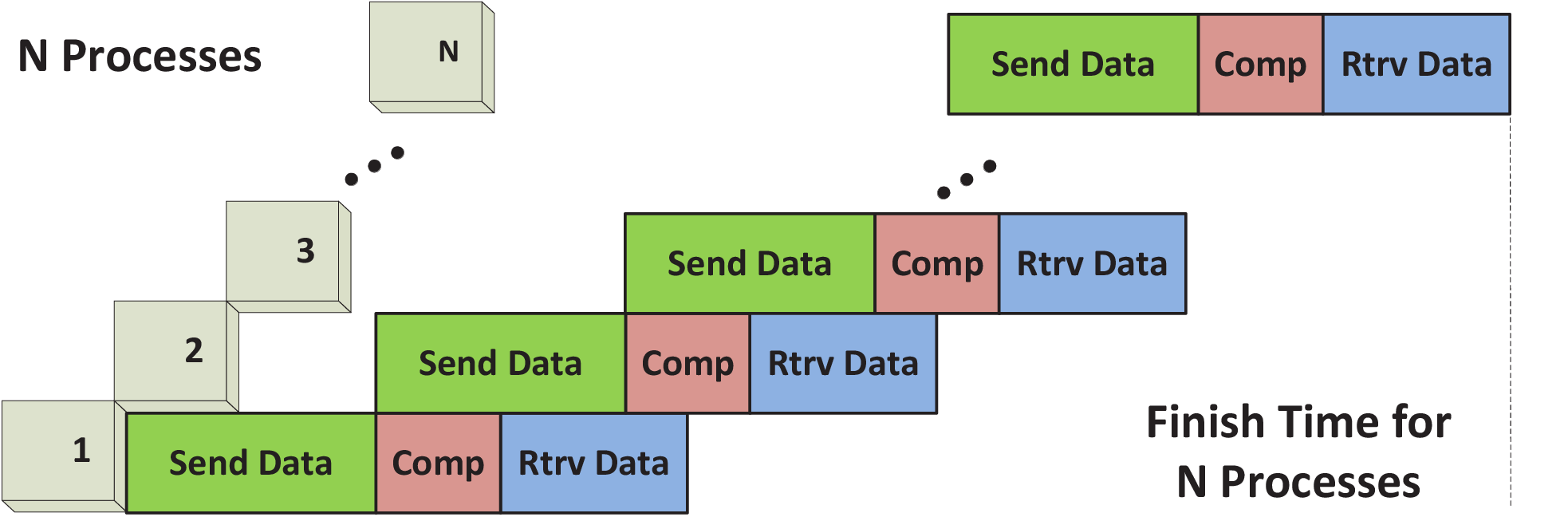}
    \caption{I/O-Intensive Kernels Under PS-2}
    \label{fig:ioi-ps2}
 \end{figure*}
\subsubsection{The GPU Execution Model Under the Optimized GPU Sharing Scheme}

To study the potential performance gain with the proposed virtualization approach, we develop a series of execution models to demonstrate the overlapping behaviors and estimate the total GPU execution time. Since the execution model is to provide an estimated performance upper bound, we assume that the GPU resource is large enough to accommodate \emph{N\textsubscript{process}} GPU kernels and also that single directional data transfers always take the full I/O bandwidth and therefore cannot be inter-overlapped. We also assume that process sequence is maintained in order since the process finishing order does not affect the total execution time. Since the daemon process is always running and has been initialized, \emph{T\textsubscript{init}} is a one-time overhead that can be hidden. Thus \emph{T\textsubscript{init}} is not considered in our analysis.

Considering the overlapping nature of two different stream programming styles, we consider two representative kernel cases: \emph{Compute-Intensive} and \emph{I/O-Intensive}.  For simplicity, the kernel is considered \emph{Compute-Intensive} when \emph{ T\textsubscript{data\_in} $\le$ T\textsubscript{comp} } and  \emph{ T\textsubscript{data\_out}  $\le$ T\textsubscript{comp} }; and defined as \emph{I/O-Intensive} when both \emph{ T\textsubscript{data\_in}} and  \emph{ T\textsubscript{data\_out}} are greater than \emph{T\textsubscript{comp} }.

Figure \ref{fig:ci-ps1} and \ref{fig:ci-ps2} show the execution and overlapping behaviors of Compute-Intensive (C-I) case written with PS-1 and PS-2 respectively. As we analysed earlier, under PS-1 shown in Figure \ref{fig:ci-ps1}, C-I kernels can overlap \emph{T\textsubscript{comp}} with both \emph{T\textsubscript{data\_in}} and \emph{T\textsubscript{comp}}. Therefore, the total execution time can be represented in Equation \eqref{eq:ci-ps1}.
\begin{align}
  \label{eq:ci-ps1}
  \emph{T\textsubscript{total\_ci\_ps1}}=\emph{N\textsubscript{process}}(\emph{T\textsubscript{data\_in}}+\emph{T\textsubscript{data\_out}})+\emph{T\textsubscript{comp}}
\end{align}
For C-I kernels programmed with PS-2, as shown in Figure \ref{fig:ci-ps2}, only I/O transfer can be inter-overlapped and overlapped with \emph{T\textsubscript{comp}}. \emph{T\textsubscript{comp}} cannot be inter-overlapped due to the implicit synchronization of \emph{T\textsubscript{data\_out}} blocking the following \emph{T\textsubscript{comp}}. Thus we are able to derive the total execution time in Equation \eqref{eq:ci-ps2}.
\begin{align}
	\label{eq:ci-ps2}
	\emph{T\textsubscript{total\_ci\_ps2}}&=\emph{T\textsubscript{data\_in}}+ \emph{N\textsubscript{process}}\emph{T\textsubscript{comp}}+\emph{T\textsubscript{data\_out}}
\end{align}
By comparing Equation \eqref{eq:ci-ps1} and \eqref{eq:ci-ps2}, we are able to see that \emph{T\textsubscript{total\_ci\_ps1}}$<$\emph{T\textsubscript{total\_ci\_ps2}}. Therefore, choosing PS-1 for C-I kernels is able to achieve better performance by concurrent kernel execution. We also note that the assumption of complete overlapping among all \emph{T\textsubscript{comp}} is merely to show the theoretical performance upper bound with the execution model.

Figure \ref{fig:ioi-ps1} and \ref{fig:ioi-ps2} demonstrate the execution scenario when I/O-Intensive (IO-I) kernels are programmed with PS-1 and PS-2 respectively. For IO-I kernels, ,the input data transfer of a subsequent kernel can only finish after the computation phase of the earlier kernel. In other words, the computation phases cannot overlap. When programmed with PS-1, the only possible overlapping is between \emph{T\textsubscript{comp}} and \emph{T\textsubscript{data\_in}}. However, since \emph{T\textsubscript{comp}} is comparatively much smaller than \emph{T\textsubscript{data\_in}}, the overlapping cannot achieve much performance improvement. The derived total execution time can be derived in Equation \eqref{eq:ioi-ps1}, which is the same as Equation \eqref{eq:ci-ps1}
\begin{align}
	\label{eq:ioi-ps1}
	\emph{T\textsubscript{total\_ioi\_ps1}}&=\emph{N\textsubscript{process}}(\emph{T\textsubscript{data\_in}}+\emph{T\textsubscript{data\_out}})+\emph{T\textsubscript{comp}}
\end{align}
Considering IO-I kernels programmed with PS-2, \emph{T\textsubscript{data\_in}} can be overlapped with both \emph{T\textsubscript{data\_out}} and \emph{T\textsubscript{comp}}, as shown in Figure \ref{fig:ioi-ps2}. We note that Figure \ref{fig:ioi-ps2}  only demonstrates the case when \emph{T\textsubscript{data\_out}} $<$ \emph{T\textsubscript{data\_in}}. By also considering the other case when \emph{T\textsubscript{data\_out}}$\ge$\emph{T\textsubscript{data\_in}}, the total execution time can be derived accordingly with both Equation \eqref{eq:ioi-ps2-1} and \eqref{eq:ioi-ps2-2}.
\begin{align}
	\label{eq:ioi-ps2-1}
	\emph{T\textsubscript{total\_ioi\_ps2}}=\emph{N\textsubscript{process}}\emph{T\textsubscript{data\_in}}+\emph{T\textsubscript{comp}}+\emph{T\textsubscript{data\_out}} \cdots	if (\emph{T\textsubscript{data\_out}} < \emph{T\textsubscript{data\_in}})
\end{align}
\begin{align}
	\label{eq:ioi-ps2-2}
	\emph{T\textsubscript{total\_ioi\_ps2}}=\emph{T\textsubscript{data\_in}}+\emph{T\textsubscript{comp}}+\emph{N\textsubscript{process}}\emph{T\textsubscript{data\_out}} 	\cdots if (\emph{T\textsubscript{data\_out}} \ge \emph{T\textsubscript{data\_in}})
\end{align}
To combine  Equation \eqref{eq:ioi-ps2-1} and \eqref{eq:ioi-ps2-2} together, we derive  Equation \eqref{eq:ioi-ps2} as the following.
\begin{align}
	\label{eq:ioi-ps2}
	\emph{T\textsubscript{total\_ioi\_ps2}}&=\emph{N\textsubscript{process}}\emph{Max(T\textsubscript{data\_in}, T\textsubscript{data\_out})}+\emph{T\textsubscript{comp}}+\emph{Min(T\textsubscript{data\_in}, T\textsubscript{data\_out}})
\end{align}

Even with the nature of no possible \emph{T\textsubscript{comp}} overlapping under PS-2, it still can achieve better performance by I/O overlapping for IO-I kernels compared with PS-1. By also comparing  Equation \eqref{eq:ioi-ps1} with \eqref{eq:ioi-ps2}, we are able to demonstrate that PS-2 brings better performance improvement specifically for the case of IO-I kernels. Therefore, our further analyses and implementation adopt PS-2 for IO-I kernels and PS-1 for C-I kernels specifically.

To estimate the potential theoretical performance gain of both C-I and IO-I kernel cases, we compare Equation \eqref{eq:ci-ps1} and \eqref{eq:ioi-ps2} with Equation \eqref{eq:seq} and therefore derive the speedups as the following two equations.
\begin{align}
  \label{eq:sp-ci}
  \emph{S\textsubscript{ci}} = \frac{\emph{T\textsubscript{total\_no\_vt}}}{\emph{T\textsubscript{total\_ci\_ps1}}} =\frac{\left(\splitfrac{\emph{N\textsubscript{process}}(\emph{T\textsubscript{init}} +\emph{T\textsubscript{data\_in}}+\emph{T\textsubscript{comp}}}
  {+\emph{T\textsubscript{data\_out}})+\left(\emph{N\textsubscript{process}}-1\right)\emph{T\textsubscript{ctx\_switch}}}\right)}{\emph{N\textsubscript{process}}\left(\emph{T\textsubscript{data\_in}}+\emph{T\textsubscript{data\_out}}\right)+\emph{T\textsubscript{comp}}}
\end{align}
\begin{align}
  \label{eq:sp-ioi}
  \emph{S\textsubscript{ioi}} = \frac{\emph{T\textsubscript{total\_no\_vt}}}{\emph{T\textsubscript{total\_ioi\_ps2}}}=\frac{\left(\splitfrac{\emph{N\textsubscript{process}}(\emph{T\textsubscript{init}} +\emph{T\textsubscript{data\_in}}+\emph{T\textsubscript{comp}}}
  { +\emph{T\textsubscript{data\_out}}) +\left(\emph{N\textsubscript{process}}-1\right)\emph{T\textsubscript{ctx\_switch}}}\right)}{\left(\splitfrac{\emph{N\textsubscript{process}}\emph{Max(T\textsubscript{data\_in}, T\textsubscript{data\_out})}}
  {+\emph{T\textsubscript{comp}}+\emph{Min(T\textsubscript{data\_in}, T\textsubscript{data\_out}})}\right)}
\end{align}
As we mentioned earlier, our main purpose with the execution model is to estimate the optimal performance upper bound with GPU virtualization. With Equation \eqref{eq:sp-ci} and \eqref{eq:sp-ioi}, the theoretical maximum speedup with GPU virtualization can be derived by considering 
\emph{N\textsubscript{process}} to infinity. If we take the limit of \eqref{eq:sp-ci} and \eqref{eq:sp-ioi}, the speedup upper bounds for both kernel cases can be derived in Equation \eqref{eq:spm-ci} and \eqref{eq:spm-ioi}.
\begin{align}
  \label{eq:spm-ci}
  \emph{S\textsubscript{max\_ci}}=\lim_{\emph{N\textsubscript{process}}\to +\infty}\emph{S\textsubscript{ci}}=\frac{\left(\splitfrac{\emph{T\textsubscript{init}} +\emph{T\textsubscript{data\_in}}+\emph{T\textsubscript{comp}}}
  {+\emph{T\textsubscript{data\_out}}+\emph{T\textsubscript{ctx\_switch}}}\right)}{\emph{T\textsubscript{data\_in}}+\emph{T\textsubscript{data\_out}}}
\end{align}
\begin{align}
  \label{eq:spm-ioi}
  \emph{S\textsubscript{max\_ioi}}=\lim_{\emph{N\textsubscript{process}}\to +\infty}\emph{S\textsubscript{ioi}} =\frac{\left(\splitfrac{\emph{T\textsubscript{init}} +\emph{T\textsubscript{data\_in}}+\emph{T\textsubscript{comp}}}
  {+\emph{T\textsubscript{data\_out}}+\emph{T\textsubscript{ctx\_switch}}}\right)}{\emph{Max(T\textsubscript{data\_in}, T\textsubscript{data\_out})}}
\end{align}
Equation \eqref{eq:spm-ci} shows that the best theoretical speedup is achieved through limiting total \emph{T\textsubscript{comp}} through concurrent kernel execution under PS-1 as well as eliminating both initialization and context-switch overhead for C-I kernels.  Equation \eqref{eq:spm-ioi} also demonstrates that I/O-I kernels can achieve the optimal performance gain with concurrent I/O under PS-2 as well as the elimination of both \emph{T\textsubscript{init}} and \emph{T\textsubscript{ctx\_switch}} overheads with our virtualization technique, the implementation details of which is to be discussed in the following.

\begin{figure*}[t]
  \begin{minipage}[b]{0.5\linewidth}    
    \centering
    \includegraphics[width= 0.9\linewidth]{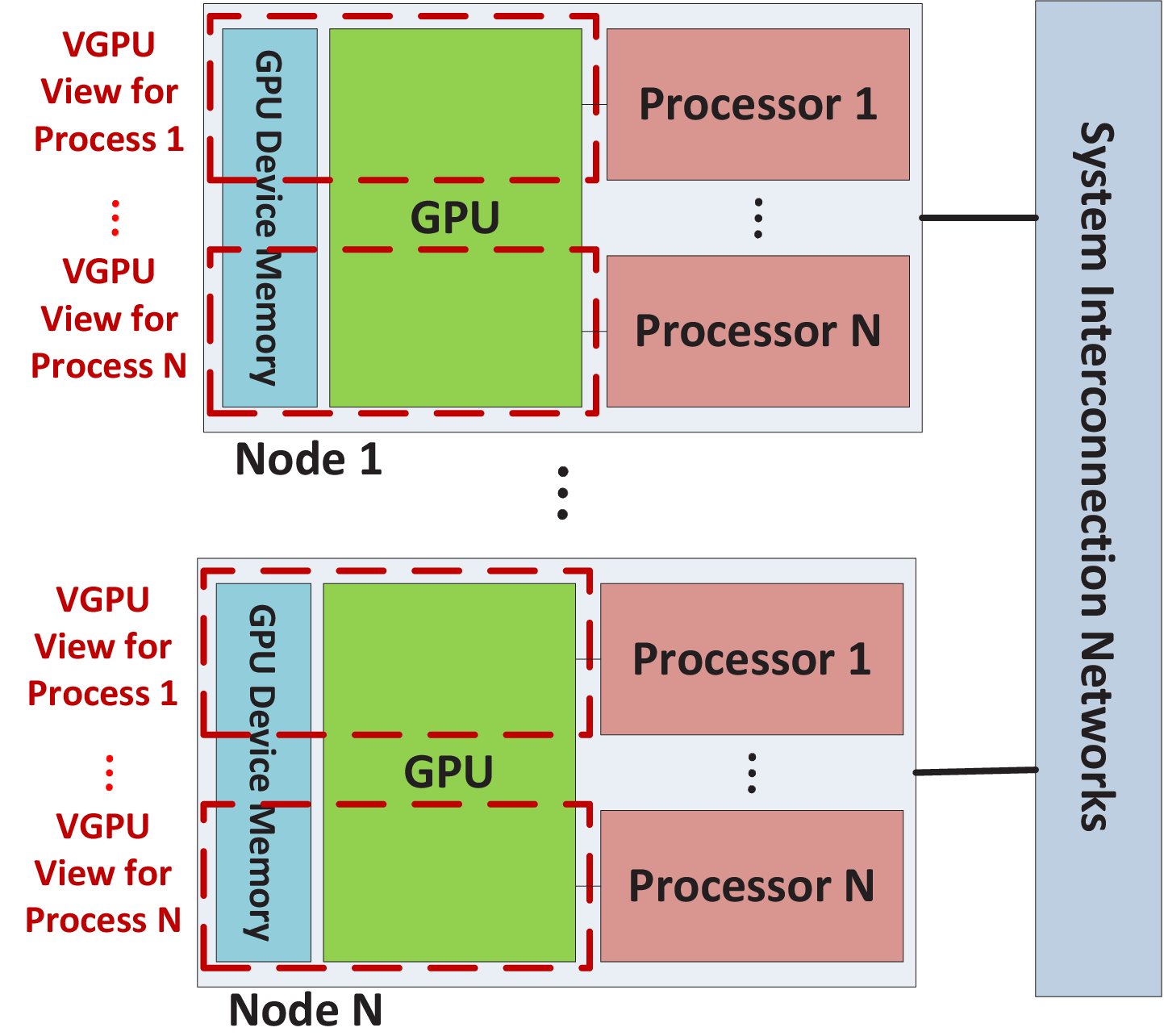}
    \caption{A Virtual SPMD View of Asymmetric Heterogeneous Cluster with GPU Virtualization}
    \label{fig:vgpu}
   \end{minipage}
  \begin{minipage}[b]{0.5\linewidth}    
    \centering
    \includegraphics[width= 0.9\linewidth]{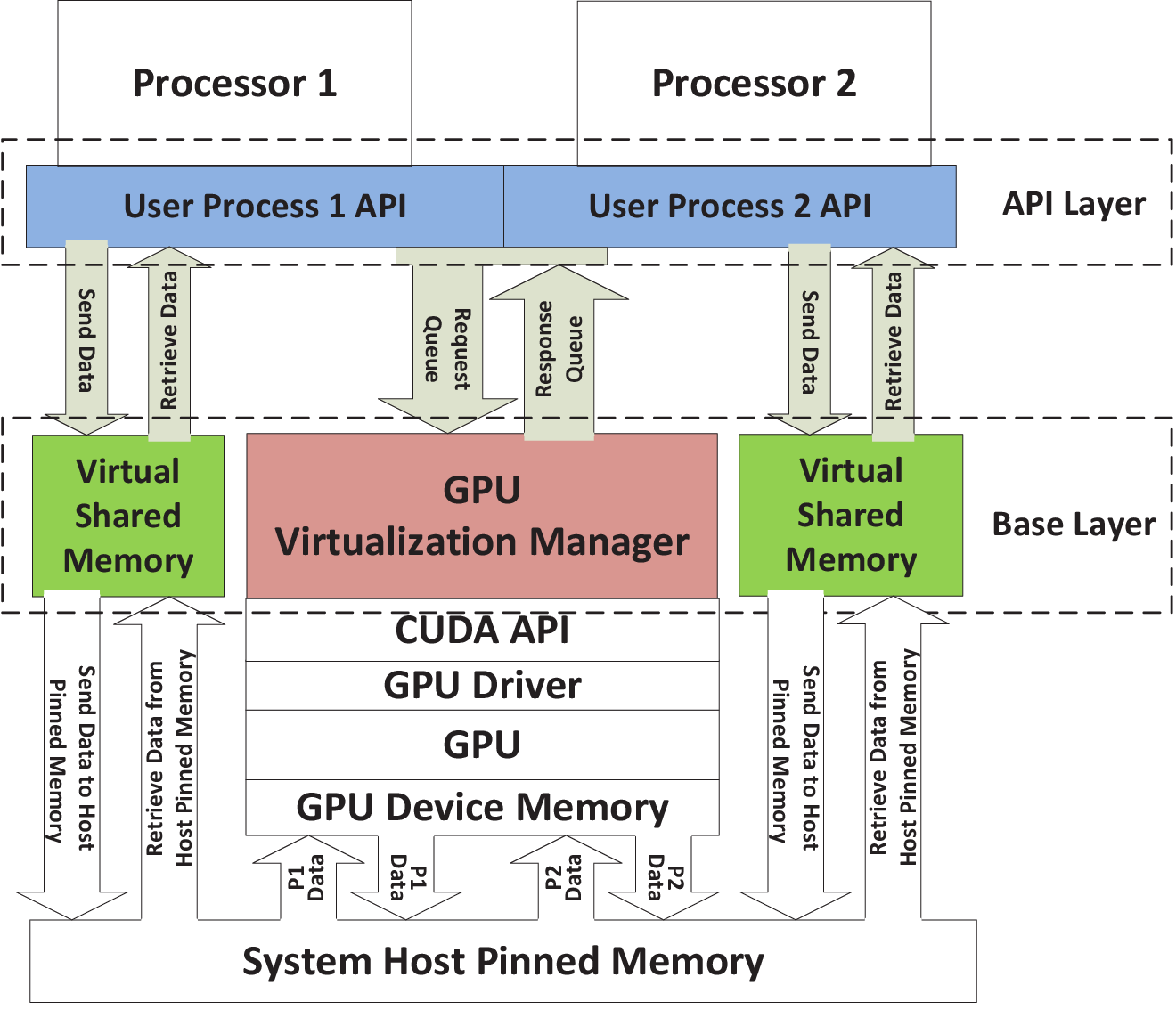}
    \caption{A Hierarchical View of the GPU Virtualization Infrastructure with Data Flow and Synchronization}
    \label{fig:infra}
   \end{minipage}
 \end{figure*}

\section{The GPU Virtualization Infrastructure}
\label{infra}

As we theoretically modeled and analyzed process-level SPMD execution parallelism and overlapping behaviors under the sharing scheme provided by GPU virtualization. In this section, we layout the GPU virtualization implementation. Figure \ref{fig:vgpu} shows a representative HPC system architecture with several nodes connected by an interconnection network. Within each node there is a heterogeneous asymmetry between the microprocessors and GPU. Under the SPMD scenario, each processor runs the same application on different data while the application has a few computationally intensive functions (GPU kernels) to be executed on the GPU. The proposed GPU virtualization infrastructure is implemented as a virtualization layer that provides a virtual view of the GPU to each processor. In other words, each of the processes running on processors sees its own Virtual GPU (VGPU) and execute its own GPU kernel and data. The deployment of the virtualization infrastructure on each computing node establishes the virtual 1 to 1 CPU/VGPU ratio and eliminates potential resource underutilization with efficient GPU sharing. 

Figure \ref{fig:infra} shows the hierarchical view of the virtualization infrastructure and data flows between layers. The base layer of the infrastructure manages the underlying GPU computing and memory resources. On top of the base layer lies the user process API layer. The base layer is consisted of the GPU Virtualization Manager (GVM), virtual shared memory space for each processor as well as request/response message queues. The API layer, which provides a virtual GPU abstraction to each of the user processes, physically handles the inter-layer communication as well as data-transfer.

In the base layer, the GVM is a run-time process responsible for initializing all virtualization resources, handling requests from processes and processing requests on the GPU device.  The initialization of the GVM creates the virtualization resources including the virtual shared memory spaces and the request/response queues. The virtual shared memory space is implemented as POSIX shared memories for each of the processes so that each process has its own virtual memory space. The virtual shared memory space handles data exchanges between processes and the GVM. Furthermore, the shared memory size is user-customizable to ensure the total size does not exceed the GPU memory size. The request/response queues are implemented as POSIX message queues to stream the process requests into the GVM and provide handshaking synchronization responses. By using streaming queues, resource contention problems are prevented. The GVM also sets request barriers to ensure that SPMD tasks from different processes can be executed in parallel. On the GPU side, when initialized, the GVM creates the necessary GPU resources including the only required GPU context and CUDA streams for each process. It also creates different memory objects for each process separately to ensure data from different processes can co-exist in the GPU memory safely. These memory objects include both GPU device memory and host pinned memory. While host pinned memory provides better I/O bandwidth, it is also required to be setup for achieving concurrent I/O and kernel execution using asynchronous streams. Moreover, the GVM takes the requested CUDA kernel functions and prepares the kernels to be executed when initialized.

\begin{figure*}[t]
 \centering
    \includegraphics[width= 1\linewidth]{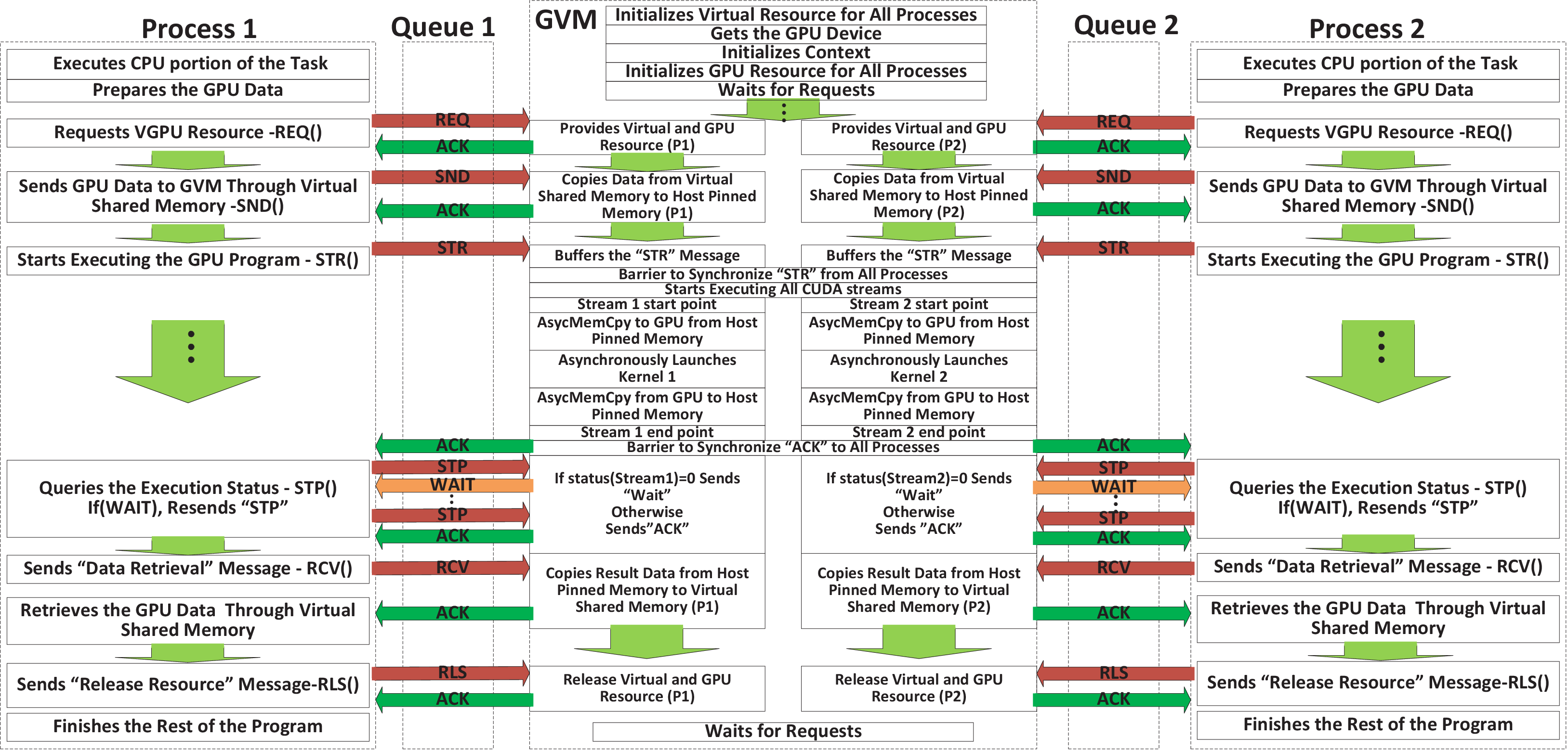}
    \caption{The Detailed Execution and Synchronization Flow of the GVM and Two User Processes}
    \label{fig:vmflow}
 \end{figure*}

The abstraction of the API layer allows the user processes to interact with the underlying run-time virtualization layer in a transparent way. While the API layer exposes the user processes (the programmers) with a virtual GPU resource view, the programmers will only need to provide the base layer with the GPU kernel function they wish to execute on their individual VGPUs. The programmer also need to take care of data exchange with the virtual shared memory, by following the procedure shown in Figure \ref{fig:vmflow} using our provided API routines such as SND()-send data routine , STR()-start execution routine and RCV()-receive data routine etc. Thus it requires very little effort to port existing GPU programs into the virtualization infrastructure. 

When an SPMD program executes under GPU virtualization, the GVM initially sets up all virtual resources associated with each process and gets ready to serve process requests. All processes accordingly send requests of VGPU resource using REQ() function while all requests are queued into the GVM to be processed. The GVM always sends the ACK signal notifying the process if the requests have been served. Each user process then continues sending data to its own share memory space and using SND() function to request GVM to process the input data from the shared memory space. Once acknowledged, the process continues to send kernel execution request through STR() followed by the query API - STP() until the ACK signal comes in (indicating that the result data is ready to be picked up from the shared memory space).  Then the user process copies the result data back from the shared memory space though RCV() routine followed by using RLS() routine to request GVM to release all the VGPU resources associated with the process itself. The detailed execution flows and interaction of the two layer are demonstrated in Figure \ref{fig:vmflow}. Here we use a two-process SPMD program example to demonstrate the interaction and synchronization between the processes and the GVM. Since requests from multiple processes come roughly simultaneously under SPMD scenario, we also set necessary execution barriers in the GVM to flush multiple CUDA streams simultaneously. We also note that different stream programming styles are used according to our previous analytical results. In other words, inside the GVM, Compute-Intensive kernels are executed with PS-1 while PS-2 is adopted by I/O-Intensive kernels for the possibly optimal concurrencies and overlapping. 

\section{Experiments and Performance Evaluation}
\label{exp}

In this section, in order to demonstrate that using the proposed GPU virtualization approach and infrastructure can provide effective resource sharing and performance gains under the SPMD execution, we conduct a series of experiments. Here we use different emulated process-level SPMD benchmarks, which launch the same benchmark program in different processes, with the affinity of each process set to a unique CPU core. We essentially conduct the experiments to compare with the same emulated process-level SPMD programs without virtualization. In other words, the performance when each process shares the GPU natively, as the performance baseline we modeled. We are mostly interested in comparing the process turnaround time, which is the time for all processes to finish executing the benchmarks after they start simultaneously. 

The experiments are conducted on our GPU computing node, which is equipped with dual Intel Xeon X5570 quad-core processors (eight cores in total) running at 2.93 GHz, 48GB system memory and the NVIDIA Tesla C2070 GPU with 6GB device memory. Tesla C2070 consists of 14 SMs running at 1.15GHz and allows maximum 16 concurrent running kernels. Both CUDA driver and SDK versions are 5.0, which run under Ubuntu 12.04 with 3.2.0-23 Linux kernel.
\begin{figure*}[t]
  \begin{minipage}[b]{0.5\linewidth}    
    \centering
    \includegraphics[width= 1\linewidth]{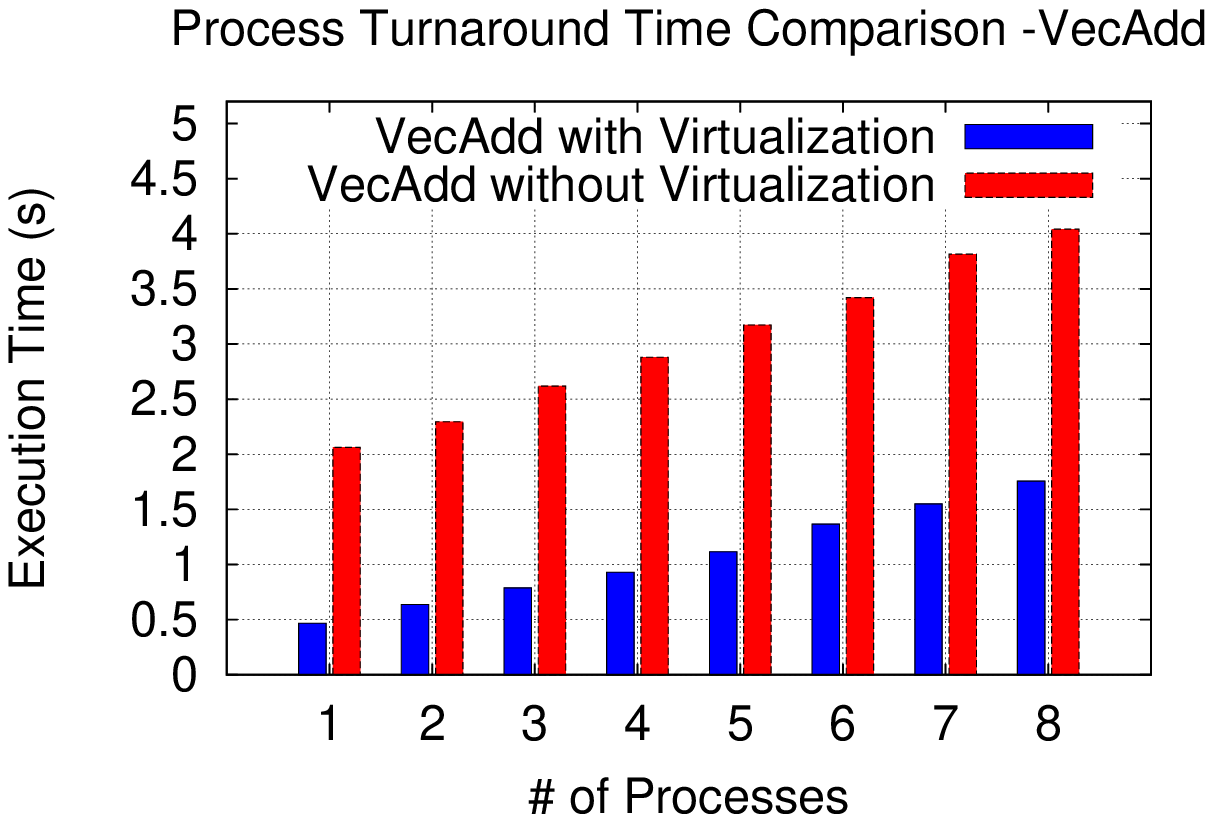}
    \caption{Process Turnaround Time Comparison: IO-I}    
    \label{fig:vecadd}
   \end{minipage}
  \begin{minipage}[b]{0.5\linewidth}    
    \centering
    \includegraphics[width= 1\linewidth]{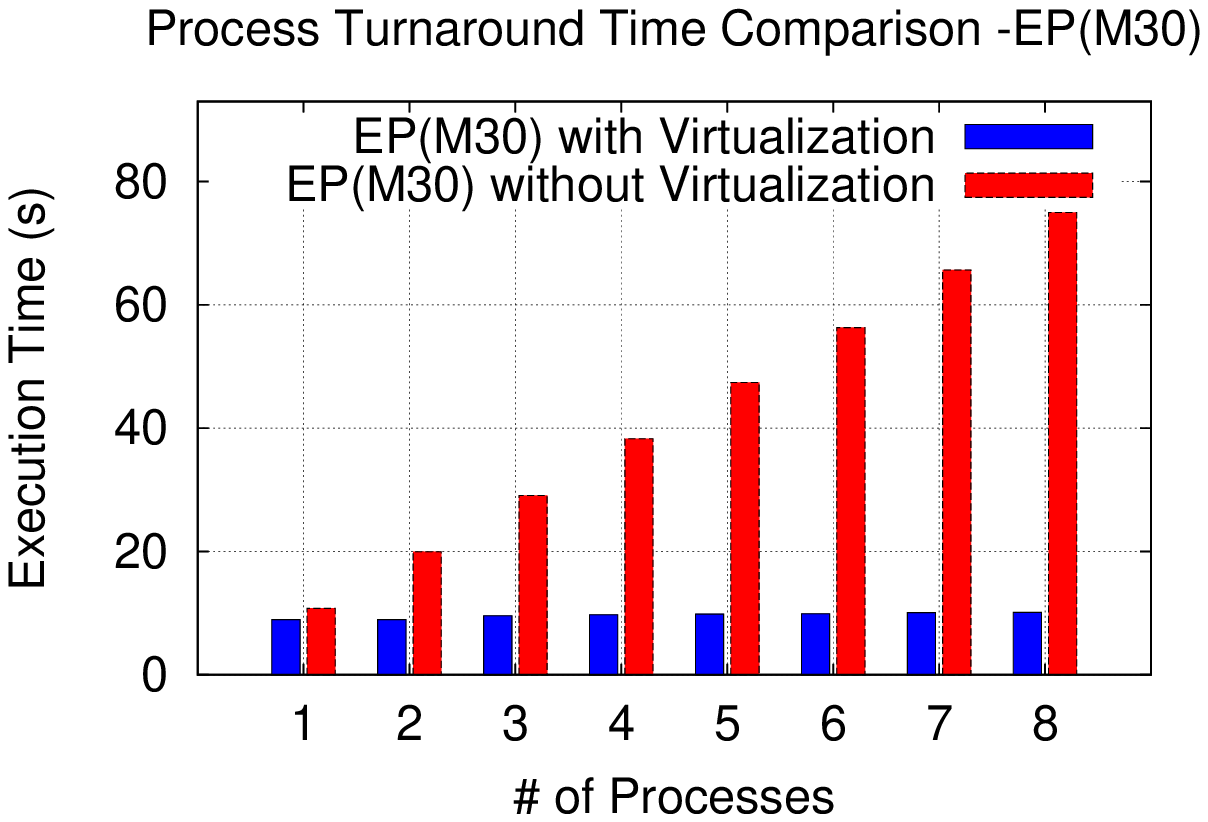}
    \caption{Process Turnaround Time Comparison: C-I}
    \label{fig:ep}
   \end{minipage}
 \end{figure*}

\begin{table*}[t]
  \caption{GPU Virtualization Benchmark Profiles}
  \label{tb:apps}
  \centering
  \begin{tabular}{@{}l@{}|l@{}|l@{}|l@{}}
	\hline
	\textbf{Benchmark} & \textbf{Problem Size} & \textbf{Grid Size} & \textbf{Class}\\
    	\hline
	NPB EP(M30) & M=30 & 4 & Compute-Intensive\\
	\hline
	Vector Addition (VecAdd) & 50M Float & 50K & I/O-Intensive\\
	\hline
	NPB EP(M24) & M=24 & 1 & Compute-Intensive\\
	\hline
	Vector Multiplication (VecMul) & 16M Float / 15 Iters & 16K & I/O-Intensive\\
	\hline
	Matrix Multiplication (MM) & 2Kx2K Matrix & 4K & Intermediate\\
	\hline
	NPB MG & S (32x32x32 / 4 Iters) & 64 & Compute-Intensive\\
	\hline
	BlackScholes (BS) & 1M Calls / 512 Iters & 480 & I/O-Intensive\\
	\hline
	NPB CG & S (NA=1400 / 15 Iters) & 8 & Compute-Intensive\\
	\hline
	Electrostatics (ES) & 100K Atoms / 25 Iters & 288 & Compute-Intensive\\
	\hline
  \end{tabular}
\end{table*}

Our previous execution models depict inter-process parallelism and overlapping under the sharing scheme achieved through GPU virtualization. As we use two kernel cases to analyze the overlapping behaviors in the execution model, here we first utilize two extreme benchmark cases (Highly Compute-Intensive and Highly I/O-Intensive) as experimental evaluation of potential overlapping behavior. The purpose is to demonstrate the different overlapping behavior for C-I and IO-I kernel cases and compare the actual performance gain with non-virtualization solution. The I/O-Intensive application we use is a very large vector addition benchmark while the Compute-Intensive benchmark is the GPU version \cite{CPE:CPE1860} of EP (Problem Size: M=30) from NAS Parallel Benchmarks (NPB) \cite{bailey1991parallel}. The EP kernel grid size is designed small merely to show the effectiveness of concurrency under virtualization, while the actual grid size decides the overlapping and concurrency extent in real applications. We list all benchmark kernel profiles used in this section in Table \ref{tb:apps}. 

Our experiment with EP(M30) and VecAdd  primarily focuses on evaluating process turnaround time by emulating a process-level parallel SPMD program for both benchmarks, while launching multiple processes with the same benchmark kernel simultaneously. As an SPMD program generally requires \emph{N\textsubscript{process}}$\le$\emph{N\textsubscript{processor}} and our computing node consists of 8 microprocessor cores, we varied the number of SPMD parallel processes from 1 to the maximum of 8. Figure \ref{fig:vecadd} and \ref{fig:ep} demonstrate the effectiveness of GPU virtualization in reducing the process turnaround time with the increasing number of processes for both C-I and IO-I cases.  For the I/O-Intensive benchmark in Figure \ref{fig:vecadd}, when the number of parallel processes increases, without virtualization, the turnaround time increase sharply due to the zero overlapping and context-switch overheads. With virtualization, the turnaround time still increases but much slowly comparatively. This is because I/O-Intensive application cannot achieve complete overlapping as explained in the model earlier, but can still partially overlap I/O as well as eliminate context-switch and initialization overheads. For the Compute-Intensive benchmark in Figure \ref{fig:ep}, with virtualization, the turnaround time increases very little with the increasing number of processes, which clearly shows that our GPU virtualization implementation can achieve the expected execution concurrencies for smaller kernels only using a portion of the GPU resource in the case of Compute-Intensive kernel. 

\begin{figure*}[t]
  \begin{minipage}[b]{0.5\linewidth}    
    \centering
    \includegraphics[width= 1\linewidth]{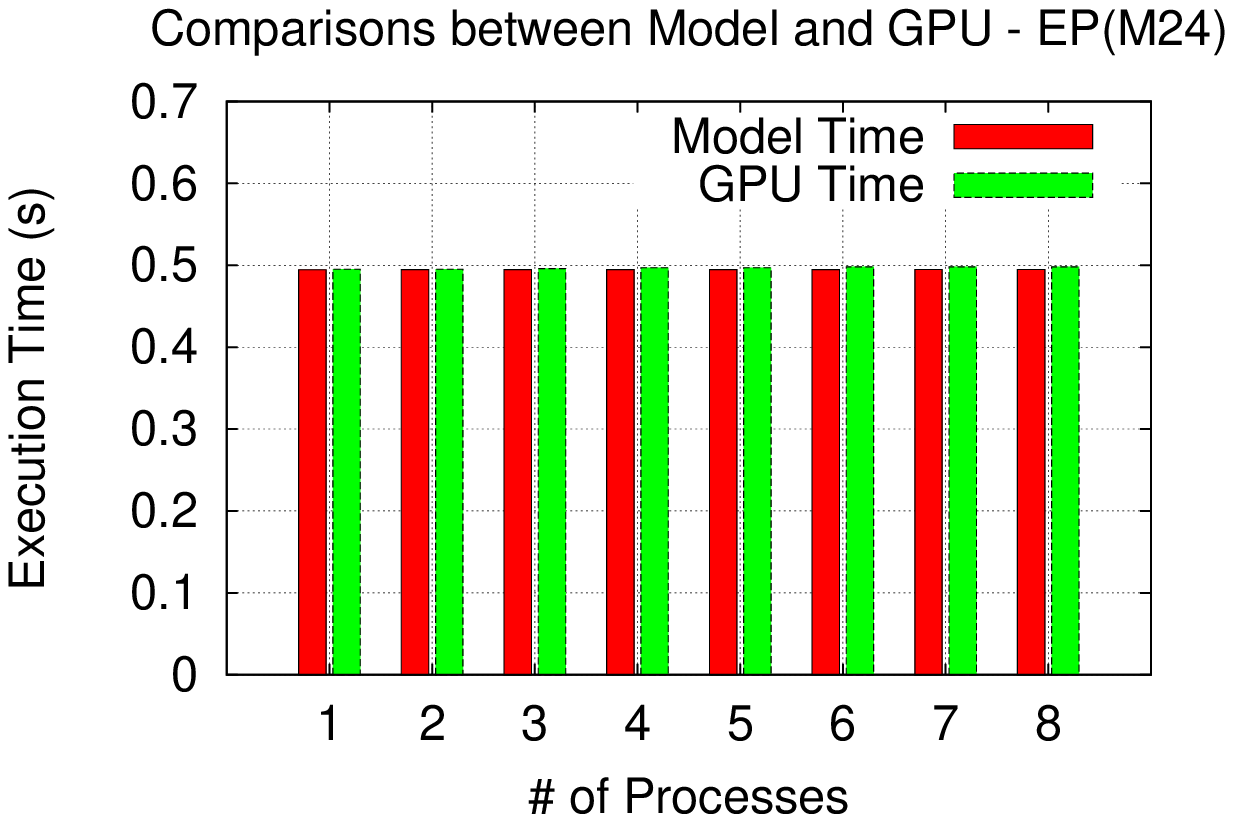}
    \caption{Execution Model Validation: C-I}    
    \label{fig:ep_model}
   \end{minipage}
   \begin{minipage}[b]{0.5\linewidth}    
    \centering
    \includegraphics[width= 1\linewidth]{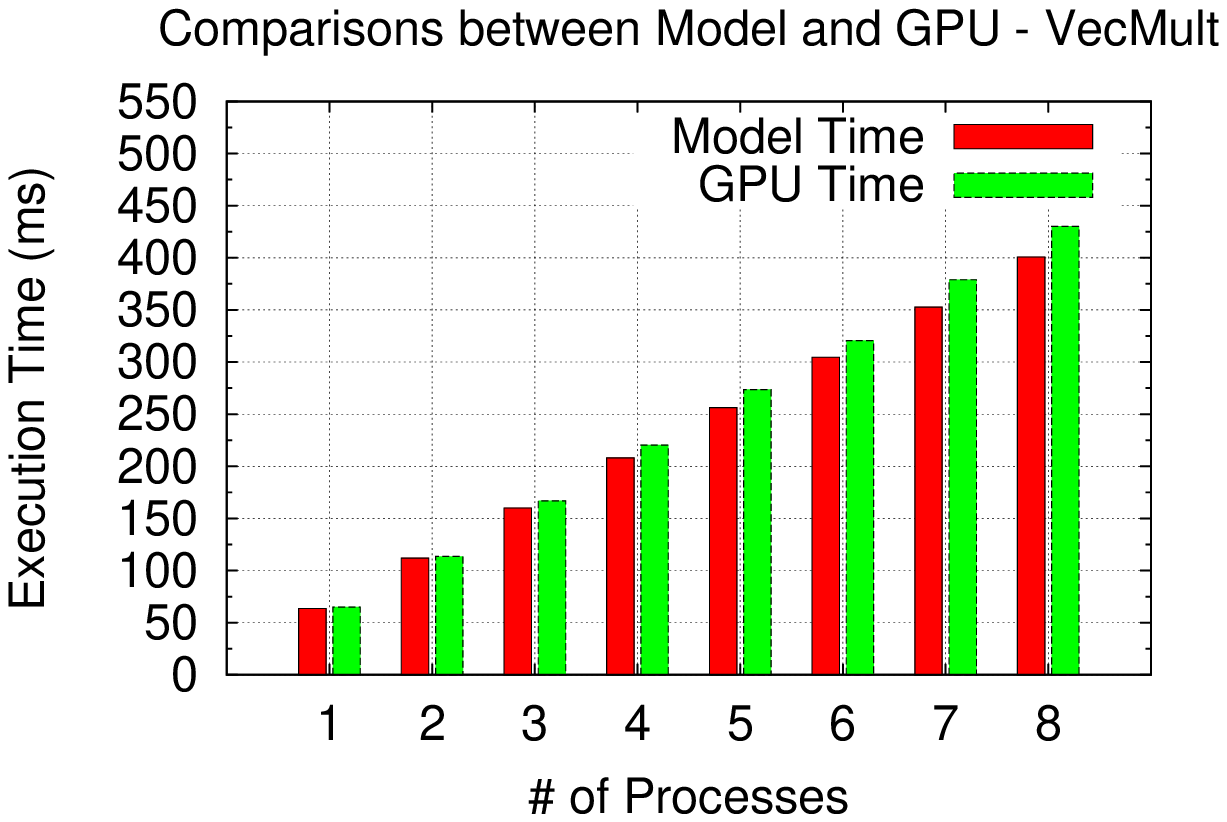}
    \caption{Execution Model Validation: IO-I}
    \label{fig:vecm_model}
   \end{minipage}
\end{figure*}
\begin{figure*}[t]
  \begin{minipage}[b]{0.5\linewidth}    
    \centering
    \includegraphics[width= 1\linewidth]{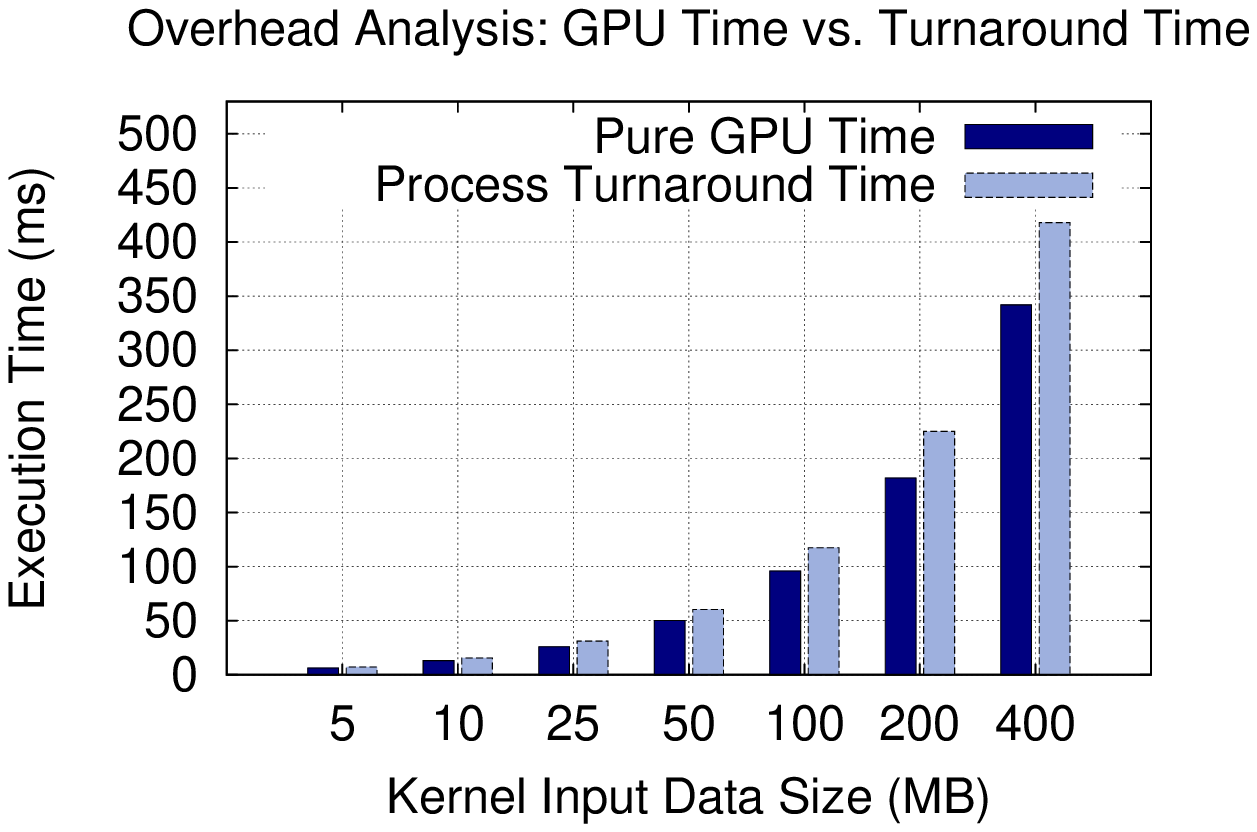}
    \caption{Virtualization Overhead}
    \label{fig:overhead}
   \end{minipage}
  \begin{minipage}[b]{0.5\linewidth}    
    \centering
    \includegraphics[width= 1\linewidth]{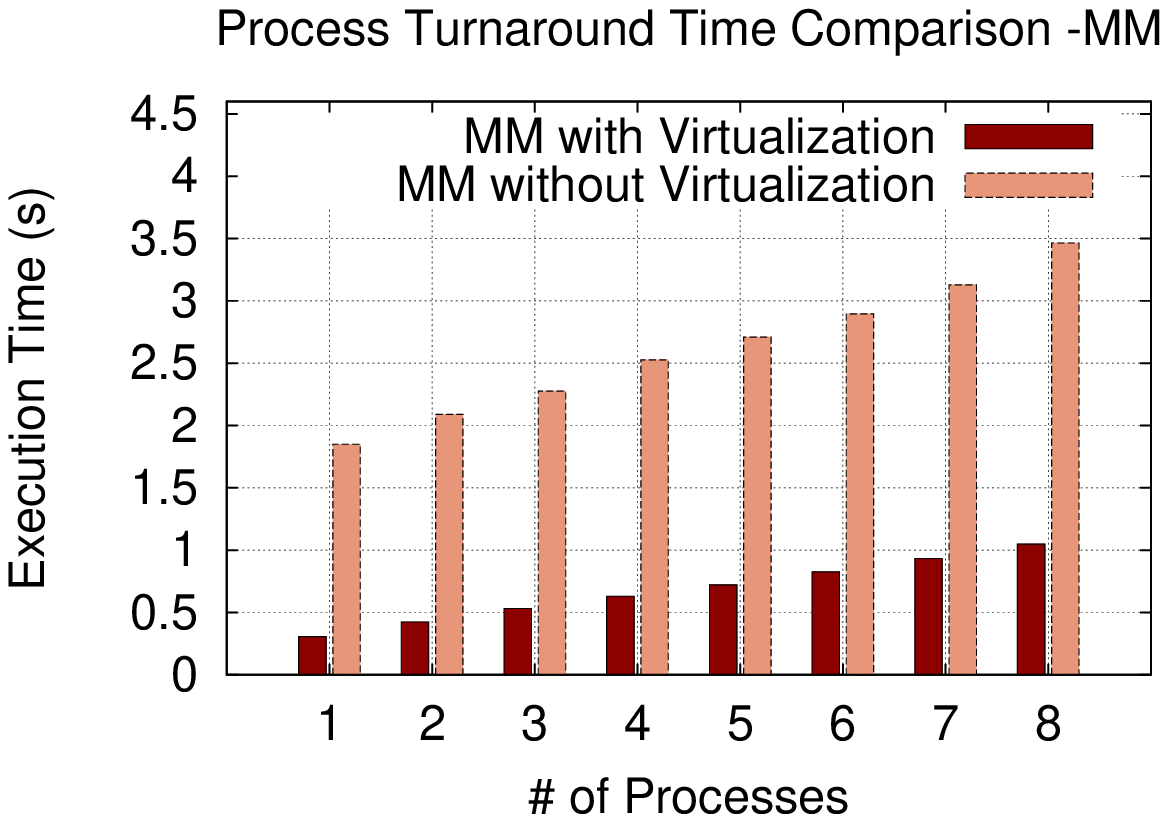}
    \caption{Performance: MM}    
    \label{fig:mm}
   \end{minipage}
  \end{figure*}
\begin{figure*}[t]
  \begin{minipage}[b]{0.5\linewidth}    
    \centering
    \includegraphics[width= 1\linewidth]{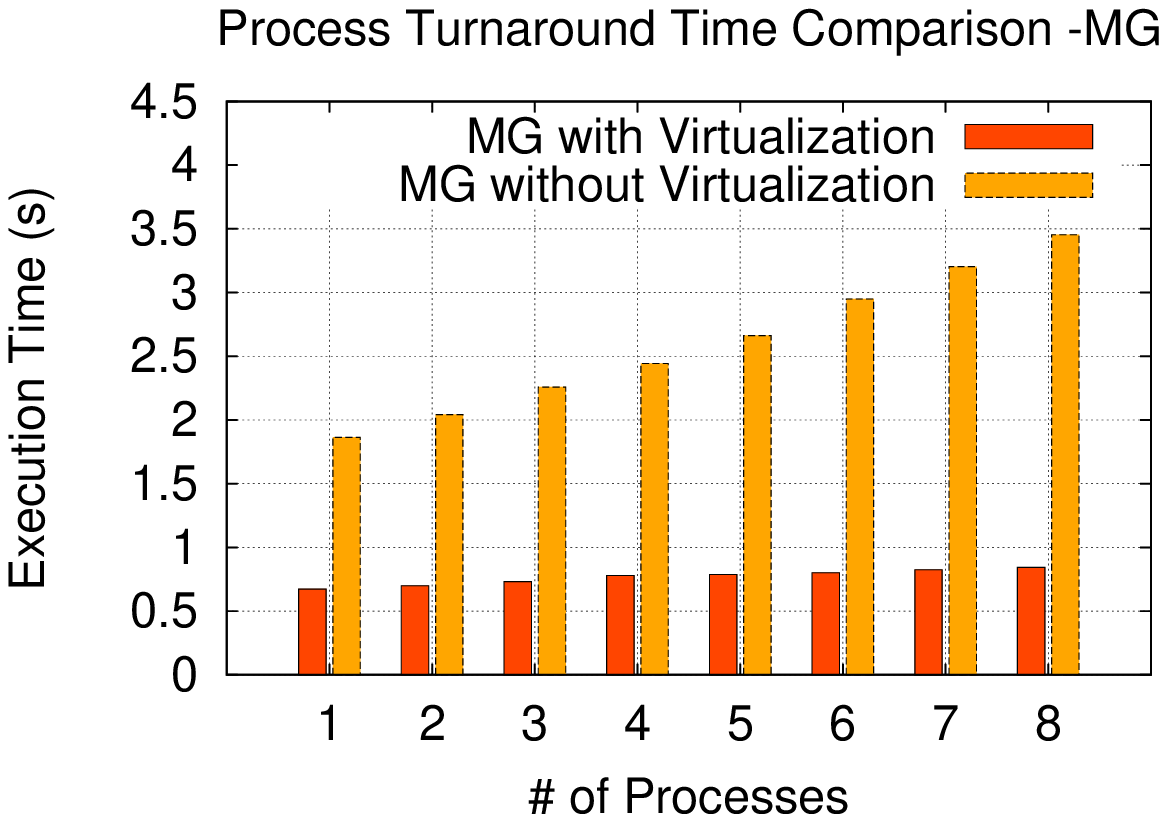}
    \caption{Performance: MG}
    \label{fig:mg}
   \end{minipage}
  \begin{minipage}[b]{0.5\linewidth}    
    \centering
    \includegraphics[width= 1\linewidth]{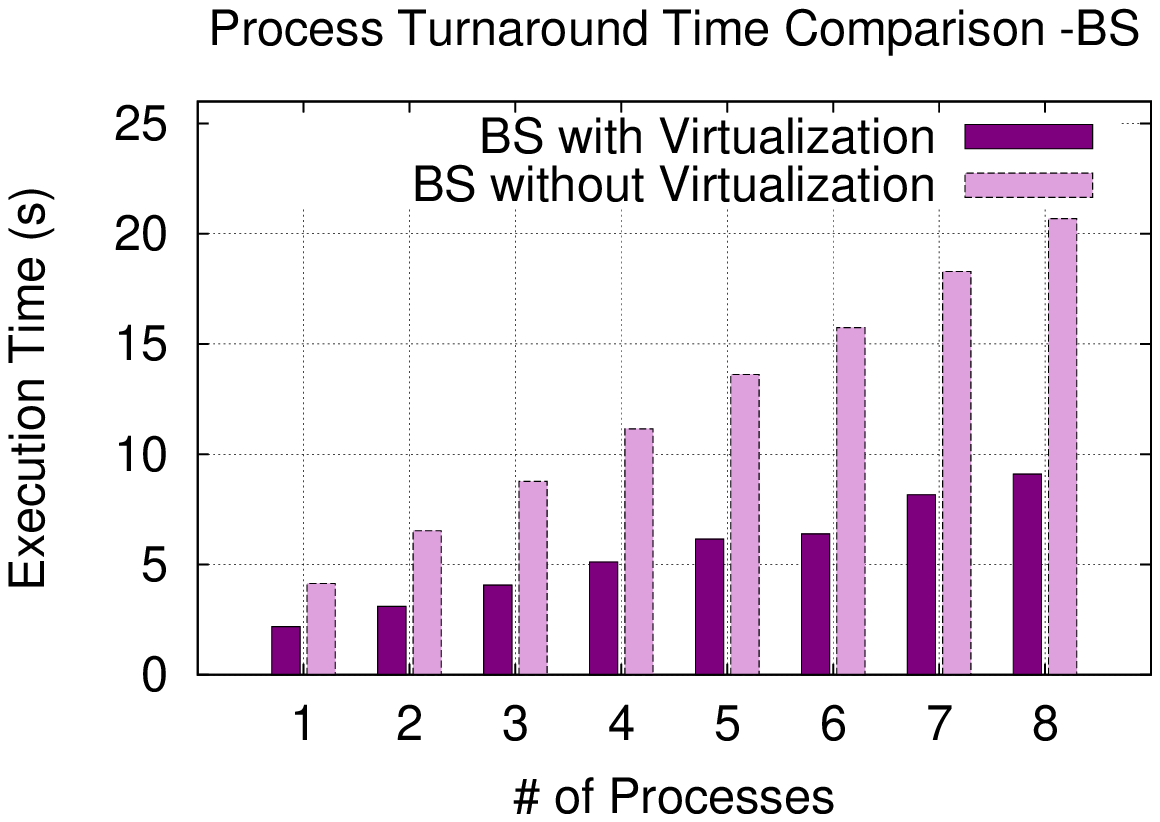}
    \caption{Performance: BS}
    \label{fig:bs}
   \end{minipage}
 \end{figure*}
\begin{figure*}[t]
  \begin{minipage}[b]{0.5\linewidth}    
    \centering
    \includegraphics[width= 1\linewidth]{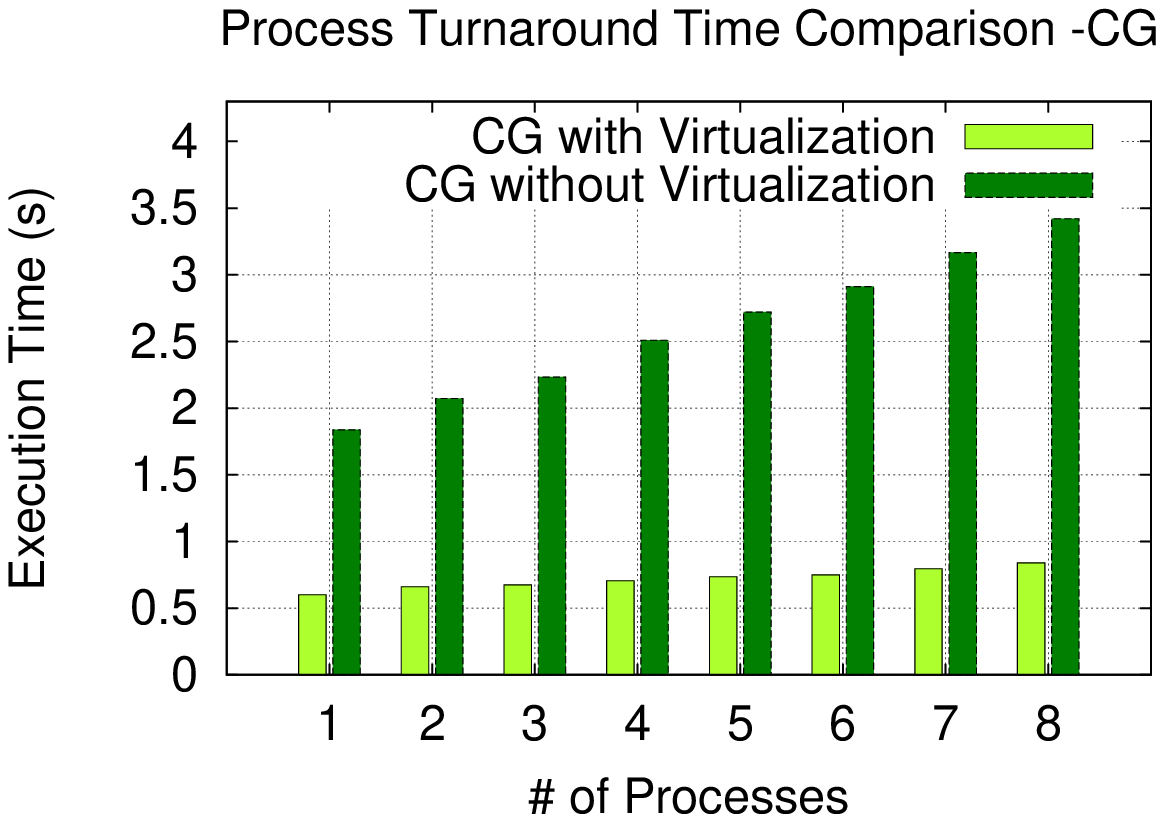}
    \caption{Performance: CG}    
    \label{fig:cg}
   \end{minipage}
   \begin{minipage}[b]{0.5\linewidth}    
    \centering
    \includegraphics[width= 1\linewidth]{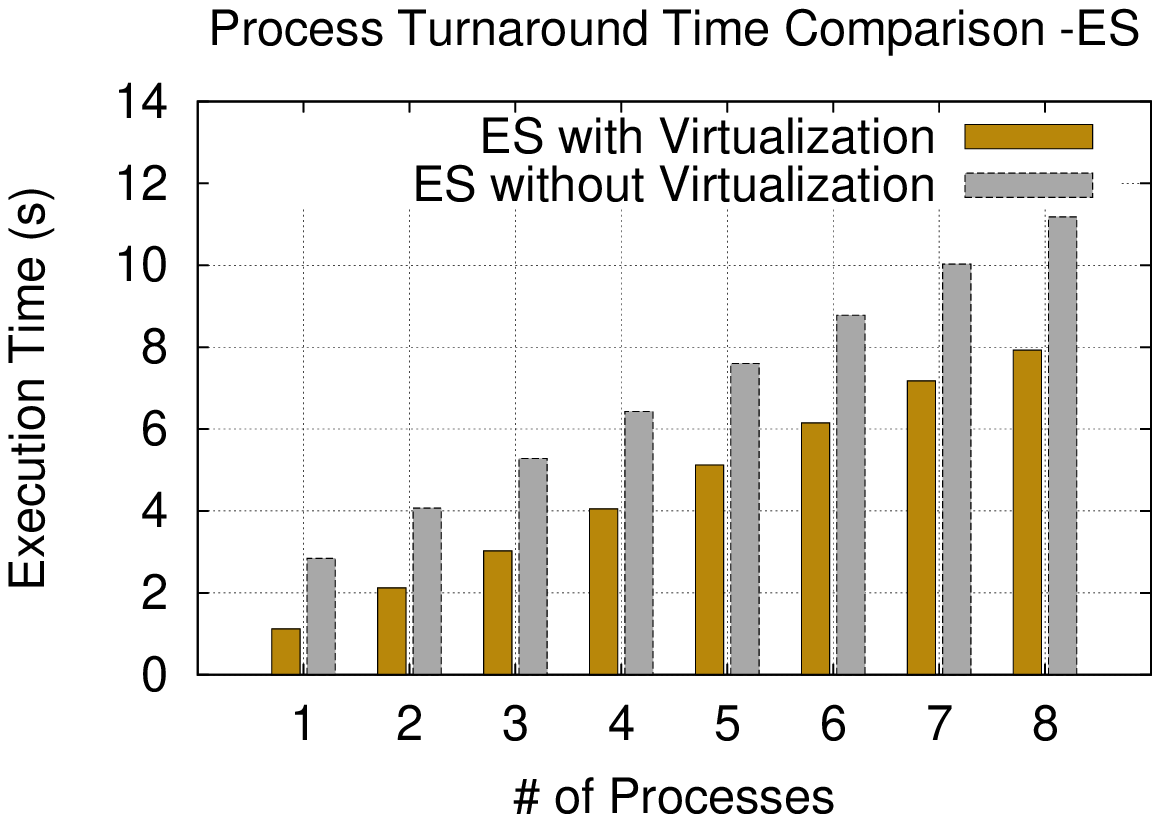}
    \caption{Performance: ES}
    \label{fig:es}
   \end{minipage}
 \end{figure*}
\begin{figure*}[t]
   \centering
   \includegraphics[width= 0.5\linewidth]{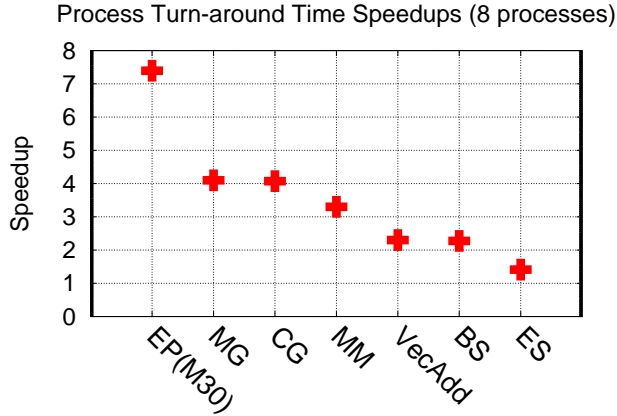}
   \caption{Virtualization Speedups}
   \label{fig:sp}
 \end{figure*}

CUDA's current concurrent kernel execution support heavily depends on kernel profiles. In other words, blocks from multiple kernels are concurrently executed on separated SMs inside GPU to achieve the concurrency when CUDA streams are used. Thus small kernels (small number of blocks) can achieve better kernel execution concurrency compared with large kernels. In the previous modeling analysis, we assume kernel execution overlapping is complete since we are focused on studying overlapping behaviors. Thus in order to verify our previous modeling analysis, we utilize EP(M24) and VecMult shown in Table \ref{tb:apps} as the benchmark kernels to verify C-I and IO-I models, respectively. For both kernels, we carry out initial profiling analysis to empirically derive \emph{T\textsubscript{data\_in}}, \emph{T\textsubscript{comp}} and \emph{T\textsubscript{data\_out}}. As both execution models are to estimate the total execution time of \emph{N\textsubscript{process}} kernels sharing the GPU under virtualization. The theoretical time can be derived using Equation \eqref{eq:ci-ps1} and \eqref{eq:ioi-ps2} respectively with the profiling results. Experimentally, we here launch the emulated SPMD kernel programs while varying \emph{N\textsubscript{process}} from 1 to 8. Instead of measuring process turnaround time, we here only measure the time all kernels spend on sharing the GPU inside the GVM of our virtualization infrastructure. Thus, we are able to avoid bringing unnecessary virtualization overheads into the model validation. Comparisons of experimental results and modeling results are shown in Figure \ref{fig:ep_model} for C-I model and in Figure \ref{fig:vecm_model} for IO-I model, both of which demonstrate accurate modeling results. We also note that for C-I model validation, utilizing EP(M24) with kernel size of one is merely to guarantee that all kernels are executed on separated SMs (up to 8 kernels in our case). In other words, complete overlapping of actual kernel computation can be achieved with kernels executed on separated SMs. The comparisons in both figures validate our previous execution model results with an average model deviation of 0.42\% for EP(M24) and 4.76\% for VecMult. 

Considering the virtualization infrastructure is an add-on layer, possible overhead can be added on top of the theoretical modeling results. As our implementation mainly uses POSIX shared memory and message queue, the vast majority overhead comes from data movement and message synchronization between the API layer and base layer. We here conduct another micro benchmark using the I/O-Intensive VecAdd benchmark with multiple data sizes. We measure the overhead by launching a single process and compare the time purely spent on the GPU in the base layer with the process turnaround time. As shown in Figure \ref{fig:overhead}, the overhead, which is the differences between the turnaround time and pure GPU time, increases with the size of data being transfered as expected. Even in the case when the data size is very large (400MB in our case), the virtualization overhead is measured around 20\%, which demonstrates that our virtualization implementation incurs comparatively low overhead, especially considering that an add-on virtualization layer generally brings much more overhead.

As a further step, we conduct several additional benchmarks to demonstrate the efficiency of the proposed virtualization approach in addressing real-life applications with different profiles. As Table \ref{tb:apps} shows, MM refers to the 2048x2048 single precision floating-point matrix multiplication. MG and CG refer to GPU versions \cite{CPE:CPE1860} of NPB \cite{bailey1991parallel} kernel MG and CG, respectively, with the problem size of Class S. Black Scholes \cite{black1973pricing} is a European option pricing benchmark used in financial area, adapted from NVIDIA's CUDA SDK. We set option prices over 512 iterations as default. Electrostatics refers to fast molecular electrostatics algorithm as a part of the molecular visualization program VMD \cite{w_vmd} and we set the problem size to be 100K atoms with 25 iterations.
By evaluating I/O and computing time ratio, we further profile the class of each benchmark as shown in Table \ref{tb:apps}. Experimentally, we emulate process-level SPMD execution of each benchmark kernel with multiple processes and compare the process turnaround time between virtualization and non-virtualization scenario. Figure \ref{fig:mm}, \ref{fig:mg}, \ref{fig:bs}, \ref{fig:cg} and \ref{fig:es} respectively compare the turnaround time with and without GPU virtualization. It is worth mentioning that the performance improvement using one process is due to the elimination of initialization overheads by the virtualization implementation, even with the add-on virtualization overhead. Since MM is profiled as intermediate and the grid size is large enough to occupy the whole GPU, it partially benefits from both I/O and kernel computing overlapping with virtualization. Both MG and CG are Compute-Intensive benchmarks and Class S problem sizes (small kernel sizes) only make MG and CG utilize partial GPU resource. Thus MG and CG can achieve more overlapping by concurrent kernel execution under virtualization. With the default problem size and a grid size of 480, a single Black Scholes benchmark can utilize full GPU resource and can hardly be concurrently executed under virtualization. Since it is also I/O-Intensive application, it is only able to achieve limited overlapping between the I/O and kernel-computing as described earlier. As for Electrostatic benchmark, since it is Compute-Intensive while the grid size of 288 making it occupy the whole GPU, the overlapping potential is small using virtualization. However, it still benefits from zero context-switch and initialization overhead due to virtualization. Therefore, within the five application benchmarks, as each achieves certain amount of performance gain through virtualization due to overlapping and elimination of overheads, MG and CG achieve better performance gains.

Figure \ref{fig:sp} summarizes an example speedup comparison scenario utilizing all available system processors (8 processes). Including EP(M30) and VecAdd as two extreme cases along with the five real-life benchmarks, all seven benchmarks we conduct achieve speedups from 1.4 to 7.4 with 8 process-level parallelism under GPU virtualization. Therefore, while all benchmarks can achieve certain amount of performance gains, the efficiency of the virtualization approach also depends on the profiles of the applications, including the I/O and computing time ratio as well the GPU resource usage. To summarize from Figure \ref{fig:sp}, small Compute-Intensive kernels can achieve the best performance improvement as EP(M30), MG and CG show. Intermediate kernels can achieve reasonable speedups with partial I/O and compute overlapping as shown from MM. I/O-Intensive kernels (BS and VecAdd) can only achieve I/O overlapping, while large Compute-Intensive kernels (ES) can overlap I/O (small portion) and very limited kernel execution. Thus they achieve relatively less performance gain. However, the elimination of context-switch and initialization overhead plus well-increased GPU utilization from GPU virtualization allow considerable speedup for application in general. In fact, our virtualization experimental results show a good agreement with the proposed analytical model, and demonstrate that our GPU virtualization implementation is an effective approach allowing multi-processes to share the GPU resource efficiently under SPMD model, while incurring comparatively low overhead.

\section{Conclusion}
\label{conc}

In this paper, we proposed a GPU virtualization approach which enables efficient sharing of GPU resources among microprocessors in heterogeneous HPC systems under SPMD execution model. In achieving the desired objective of making each microprocessors effectively utilize shared resources, we investigated the concurrency and overlapping potentials that can be exploited on the GPU device-level. We also analyzed the performance and overheads of direct GPU access and sharing from multiple microprocessors as a comparison baseline. We further provided an analytical execution model as a theoretical performance estimate of our proposed virtualization approach. The analytical model also provided us with better understanding of the methodologies in implementing our virtualization concept. Based on these concepts and analyses, we implemented our virtualization infrastructure as a run-time layer running in the user space of the OS. The virtualization layer manages requests from all microprocessors and provides necessary GPU resources to the microprocessors. It also exposes a VGPU view to all the microprocessors as if each microprocessor has its own GPU resource. Inside the virtualization layer, we managed to eliminate unnecessary overheads and achieve possible overlapping and concurrency of executions. In the experiments, we utilized our NVIDIA Fermi GPU computing node as the test bed. We used initial I/O-Intensive and Compute-Intensive benchmarks as well as application benchmarks with multiple folds of analyses in our experiments. Our experimental results showed that we were able to achieve considerable performance gains in terms of speedups with our virtualization infrastructure with low overhead. The experimental results also demonstrate an agreement with our theoretical analysis. Proposed as a solution for microprocessor resource underutilization by providing a virtual SPMD execution scenario, our approach proves to be effective and efficient and can be deployed to any heterogeneous GPU clusters with imbalanced CPU/GPU resources. 



%
%


\begin{thebibliography}{37}
\expandafter\ifx\csname natexlab\endcsname\relax\def\natexlab#1{#1}\fi
\providecommand{\url}[1]{\texttt{#1}}
\providecommand{\href}[2]{#2}
\providecommand{\path}[1]{#1}
\providecommand{\DOIprefix}{doi:}
\providecommand{\ArXivprefix}{arXiv:}
\providecommand{\URLprefix}{URL: }
\providecommand{\Pubmedprefix}{pmid:}
\providecommand{\doi}[1]{\href{http://dx.doi.org/#1}{\path{#1}}}
\providecommand{\Pubmed}[1]{\href{pmid:#1}{\path{#1}}}
\providecommand{\bibinfo}[2]{#2}
\ifx\xfnm\relax \def\xfnm[#1]{\unskip,\space#1}\fi
\bibitem[{Li et~al.(2011)Li, Narayana, El-Araby, and El-Ghazawi}]{c_vgpu}
\bibinfo{author}{T.~Li}, \bibinfo{author}{V.~K. Narayana},
  \bibinfo{author}{E.~El-Araby}, \bibinfo{author}{T.~El-Ghazawi},
\newblock \bibinfo{title}{{GPU} resource sharing and virtualization on high
  performance computing systems},
\newblock in: \bibinfo{booktitle}{Parallel Processing (ICPP), 2011
  International Conference on}, \bibinfo{organization}{IEEE},
  \bibinfo{year}{2011}, pp. \bibinfo{pages}{733--742}.
  \DOIprefix\doi{10.1109/ICPP.2011.88}.
\bibitem[{wp_(2013)}]{wp_gpgpu}
\bibinfo{title}{{GPGPU} webpage},
\newblock \bibinfo{year}{2013}.
\bibitem[{{NVIDIA Corp}.(2010)}]{wp_tesla}
\bibinfo{author}{{NVIDIA Corp}.}, \bibinfo{title}{Tesla GPU Computing
  Brochure}, \bibinfo{year}{2010}.
\bibitem[{AMD(2013)}]{wp_amd13}
\bibinfo{author}{AMD}, \bibinfo{title}{{AMD FirePro S10000 Datasheet}},
  \bibinfo{year}{2013}.
\bibitem[{Fan et~al.(2004)Fan, Qiu, Kaufman, and Yoakum-Stover}]{fan2004gpu}
\bibinfo{author}{Z.~Fan}, \bibinfo{author}{F.~Qiu},
  \bibinfo{author}{A.~Kaufman}, \bibinfo{author}{S.~Yoakum-Stover},
\newblock \bibinfo{title}{{GPU cluster for high performance computing}},
\newblock in: \bibinfo{booktitle}{Proceedings of the 2004 ACM/IEEE conference
  on Supercomputing}, \bibinfo{organization}{IEEE Computer Society},
  \bibinfo{year}{2004}, p.~\bibinfo{pages}{47}.
\bibitem[{Kindratenko et~al.(2009)Kindratenko, Enos, Shi, Showerman, Arnold,
  Stone, Phillips, and Hwu}]{kindratenko2009gpu}
\bibinfo{author}{V.~Kindratenko}, \bibinfo{author}{J.~Enos},
  \bibinfo{author}{G.~Shi}, \bibinfo{author}{M.~Showerman},
  \bibinfo{author}{G.~Arnold}, \bibinfo{author}{J.~Stone},
  \bibinfo{author}{J.~Phillips}, \bibinfo{author}{W.~Hwu},
\newblock \bibinfo{title}{{GPU clusters for high-performance computing}},
\newblock in: \bibinfo{booktitle}{Cluster Computing and Workshops, 2009.
  CLUSTER'09. IEEE International Conference on}, \bibinfo{organization}{IEEE},
  \bibinfo{year}{2009}, pp. \bibinfo{pages}{1--8}.
\bibitem[{{Cray Inc.}(2012)}]{m_crayxk7}
\bibinfo{author}{{Cray Inc.}}, \bibinfo{title}{Cray XK7 Brochure},
  \bibinfo{year}{2012}.
\bibitem[{{SGI Corp.}(2012)}]{m_sgi}
\bibinfo{author}{{SGI Corp.}}, \bibinfo{title}{{SGI GPU Compute Solutions}},
  \bibinfo{year}{2012}.
\bibitem[{w_t(2013{\natexlab{a}})}]{w_titan}
\bibinfo{title}{{Titan Webpage in Oak Ridge National Lab}},
\newblock \bibinfo{year}{2013}{\natexlab{a}}.
\bibitem[{w_t(2013{\natexlab{b}})}]{w_top500}
\bibinfo{title}{{Top 500 Supercomputer Sites Webpage}},
\newblock \bibinfo{year}{2013}{\natexlab{b}}.
\bibitem[{Darema(2001)}]{darema2001spmd}
\bibinfo{author}{F.~Darema},
\newblock \bibinfo{title}{{The SPMD model: Past, present and future}},
\newblock \bibinfo{journal}{Recent Advances in Parallel Virtual Machine and
  Message Passing Interface}  (\bibinfo{year}{2001}) \bibinfo{pages}{1--1}.
\bibitem[{Gropp et~al.(1999)Gropp, Lusk, and Skjellum}]{gropp1999using}
\bibinfo{author}{W.~Gropp}, \bibinfo{author}{E.~Lusk},
  \bibinfo{author}{A.~Skjellum}, \bibinfo{title}{Using {MPI}: portable parallel
  programming with the message passing interface}, volume~\bibinfo{volume}{1},
  \bibinfo{publisher}{MIT press}, \bibinfo{year}{1999}.
\bibitem[{w_n(2013)}]{w_nscc}
\bibinfo{title}{{China National Supercomputer Center in Tianjin Webpage}},
\newblock \bibinfo{year}{2013}.
\bibitem[{w_s(2013)}]{w_sugon}
\bibinfo{title}{{Nebulae Specification in Sugon Webpage}},
\newblock \bibinfo{year}{2013}.
\bibitem[{{GSIC, Tokyo Institute of Technology}.(2011)}]{m_tsubame}
\bibinfo{author}{{GSIC, Tokyo Institute of Technology}.},
  \bibinfo{title}{{Tsubame Hardware Software Specifications}},
  \bibinfo{year}{2011}.
\bibitem[{{Advanced Micro Devices, Inc.}(2011)}]{wp_amd}
\bibinfo{author}{{Advanced Micro Devices, Inc.}}, \bibinfo{title}{{AMD
  Firestream 9350 Datasheet}}, \bibinfo{year}{2011}.
\bibitem[{{NVIDIA Corp.}(2009)}]{wp_fermi}
\bibinfo{author}{{NVIDIA Corp.}}, \bibinfo{title}{NVIDIA's Next Generation CUDA
  Compute Architecture: Fermi}, \bibinfo{year}{2009}. \bibinfo{note}{{Ver}.
  1.1}.
\bibitem[{{NVIDIA Corp}.(2013)}]{m_cuda}
\bibinfo{author}{{NVIDIA Corp}.}, \bibinfo{title}{NVIDIA CUDA C-Programming
  Guide V5.5}, \bibinfo{year}{2013}.
\bibitem[{Gupta et~al.(2009)Gupta, Gavrilovska, Schwan, Kharche, Tolia, Talwar,
  and Ranganathan}]{gupta2009gvim}
\bibinfo{author}{V.~Gupta}, \bibinfo{author}{A.~Gavrilovska},
  \bibinfo{author}{K.~Schwan}, \bibinfo{author}{H.~Kharche},
  \bibinfo{author}{N.~Tolia}, \bibinfo{author}{V.~Talwar},
  \bibinfo{author}{P.~Ranganathan},
\newblock \bibinfo{title}{{GViM: GPU-accelerated Virtual Machines}},
\newblock in: \bibinfo{booktitle}{Proceedings of the 3rd ACM Workshop on
  System-level Virtualization for High Performance Computing},
  \bibinfo{organization}{ACM}, \bibinfo{year}{2009}, pp.
  \bibinfo{pages}{17--24}.
\bibitem[{Shi et~al.(2009)Shi, Chen, and Sun}]{shi2009vcuda}
\bibinfo{author}{L.~Shi}, \bibinfo{author}{H.~Chen}, \bibinfo{author}{J.~Sun},
\newblock \bibinfo{title}{{vCUDA: GPU} accelerated high performance computing
  in virtual machines},
\newblock in: \bibinfo{booktitle}{Parallel \& Distributed Processing, 2009.
  IPDPS 2009. IEEE International Symposium on}, \bibinfo{organization}{IEEE},
  \bibinfo{year}{2009}, pp. \bibinfo{pages}{1--11}.
\bibitem[{Giunta et~al.(2010)Giunta, Montella, Agrillo, and
  Coviello}]{giunta2010gpgpu}
\bibinfo{author}{G.~Giunta}, \bibinfo{author}{R.~Montella},
  \bibinfo{author}{G.~Agrillo}, \bibinfo{author}{G.~Coviello},
\newblock \bibinfo{title}{{A GPGPU transparent virtualization component for
  high performance computing clouds}},
\newblock \bibinfo{journal}{Euro-Par 2010-Parallel Processing}
  (\bibinfo{year}{2010}) \bibinfo{pages}{379--391}.
\bibitem[{Duato et~al.(2010)Duato, Igual, Mayo, Pe{\~n}a, Quintana-Ort{\'\i},
  and Silla}]{duato2010efficient}
\bibinfo{author}{J.~Duato}, \bibinfo{author}{F.~Igual},
  \bibinfo{author}{R.~Mayo}, \bibinfo{author}{A.~Pe{\~n}a},
  \bibinfo{author}{E.~Quintana-Ort{\'\i}}, \bibinfo{author}{F.~Silla},
\newblock \bibinfo{title}{An efficient implementation of {GPU} virtualization
  in high performance clusters},
\newblock in: \bibinfo{booktitle}{Euro-Par 2009--Parallel Processing
  Workshops}, \bibinfo{organization}{Springer}, \bibinfo{year}{2010}, pp.
  \bibinfo{pages}{385--394}.
\bibitem[{Guevara et~al.(2009)Guevara, Gregg, Hazelwood, and
  Skadron}]{guevara2009enabling}
\bibinfo{author}{M.~Guevara}, \bibinfo{author}{C.~Gregg},
  \bibinfo{author}{K.~Hazelwood}, \bibinfo{author}{K.~Skadron},
\newblock \bibinfo{title}{Enabling task parallelism in the {CUDA} scheduler},
\newblock in: \bibinfo{booktitle}{Proceedings of the Workshop on Programming
  Models for Emerging Architectures (PMEA)}, \bibinfo{organization}{Citeseer},
  \bibinfo{year}{2009}.
\bibitem[{Saba and Mangharam(2010)}]{saba2010anytime}
\bibinfo{author}{A.~Saba}, \bibinfo{author}{R.~Mangharam},
\newblock \bibinfo{title}{{Anytime Algorithms for GPU Architectures}},
\newblock \bibinfo{journal}{AVICPS 2010}  (\bibinfo{year}{2010})
  \bibinfo{pages}{31}.
\bibitem[{Li et~al.(2011)Li, Narayana, and El-Ghazawi}]{c_gpusch}
\bibinfo{author}{T.~Li}, \bibinfo{author}{V.~K. Narayana},
  \bibinfo{author}{T.~El-Ghazawi},
\newblock \bibinfo{title}{A static task scheduling framework for independent
  tasks accelerated using a shared graphics processing unit},
\newblock in: \bibinfo{booktitle}{Parallel and Distributed Systems (ICPADS),
  2011 IEEE 17th International Conference on}, \bibinfo{organization}{IEEE},
  \bibinfo{year}{2011}, pp. \bibinfo{pages}{88--95}.
  \DOIprefix\doi{10.1109/ICPADS.2011.13}.
\bibitem[{Li et~al.(2014)Li, Narayana, and El-Ghazawi}]{cf14}
\bibinfo{author}{T.~Li}, \bibinfo{author}{V.~K. Narayana},
  \bibinfo{author}{T.~El-Ghazawi},
\newblock \bibinfo{title}{Symbiotic scheduling of concurrent {GPU} kernels for
  performance and energy optimizations},
\newblock in: \bibinfo{booktitle}{Proceedings of the 11th ACM Conference on
  Computing Frontiers}, CF '14, \bibinfo{publisher}{ACM}, \bibinfo{address}{New
  York, NY, USA}, \bibinfo{year}{2014}, pp. \bibinfo{pages}{36:1--36:10}.
  \URLprefix \url{http://doi.acm.org/10.1145/2597917.2597925}.
  \DOIprefix\doi{10.1145/2597917.2597925}.

\bibitem[{Li et~al.(2015)Li, Narayana, and El-Ghazawi}]{icpads15}
\bibinfo{author}{T.~Li}, \bibinfo{author}{V.~K. Narayana},
  \bibinfo{author}{T.~El-Ghazawi},
\newblock \bibinfo{title}{A power-aware symbiotic scheduling algorithm for
  concurrent gpu kernels},
\newblock in: \bibinfo{booktitle}{The 21st IEEE International Conference on
  Parallel and Distributed Systems (ICPADS 2015)},
  \bibinfo{organization}{IEEE}, \bibinfo{year}{2015}.
\bibitem[{Ravi et~al.(2011)Ravi, Becchi, Agrawal, and
  Chakradhar}]{ravi2011supporting}
\bibinfo{author}{V.~Ravi}, \bibinfo{author}{M.~Becchi},
  \bibinfo{author}{G.~Agrawal}, \bibinfo{author}{S.~Chakradhar},
\newblock \bibinfo{title}{{Supporting GPU sharing in cloud environments with a
  transparent runtime consolidation framework}},
\newblock in: \bibinfo{booktitle}{the International Symposium on
  High-Performance Parallel and Distributed Computing}, \bibinfo{year}{2011}.
\bibitem[{Peters et~al.(2010)Peters, Koper, and
  Luttenberger}]{peters2010efficiently}
\bibinfo{author}{H.~Peters}, \bibinfo{author}{M.~Koper},
  \bibinfo{author}{N.~Luttenberger},
\newblock \bibinfo{title}{{Efficiently using a CUDA-enabled GPU as shared
  resource}},
\newblock in: \bibinfo{booktitle}{Computer and Information Technology (CIT),
  2010 IEEE 10th International Conference on}, \bibinfo{organization}{IEEE},
  \bibinfo{year}{2010}, pp. \bibinfo{pages}{1122--1127}.
\bibitem[{wp_(2013)}]{wp_sgpu}
\bibinfo{title}{{S\_GPU} project webpage},
\newblock \bibinfo{year}{2013}.

\bibitem[{Li et~al.(2012)Li, Narayana, and El-Ghazawi}]{cf12}
\bibinfo{author}{T.~Li}, \bibinfo{author}{V.~K. Narayana},
  \bibinfo{author}{T.~El-Ghazawi},
\newblock \bibinfo{title}{Accelerated high-performance computing through
  efficient multi-process {GPU} resource sharing},
\newblock in: \bibinfo{booktitle}{Proceedings of the 9th Conference on
  Computing Frontiers}, CF '12, \bibinfo{publisher}{ACM}, \bibinfo{address}{New
  York, NY, USA}, \bibinfo{year}{2012}, pp. \bibinfo{pages}{269--272}.
  \URLprefix \url{http://doi.acm.org/10.1145/2212908.2212950}.
  \DOIprefix\doi{10.1145/2212908.2212950}.
\bibitem[{Li et~al.(2013)Li, Narayana, and El-Ghazawi}]{computers}
\bibinfo{author}{T.~Li}, \bibinfo{author}{V.~K. Narayana},
  \bibinfo{author}{T.~El-Ghazawi},
\newblock \bibinfo{title}{Exploring graphics processing unit ({GPU}) resource
  sharing efficiency for high performance computing},
\newblock \bibinfo{journal}{Computers} \bibinfo{volume}{2}
  (\bibinfo{year}{2013}) \bibinfo{pages}{176--214}. \URLprefix
  \url{http://www.mdpi.com/2073-431X/2/4/176}.
  \DOIprefix\doi{10.3390/computers2040176}.

\bibitem[{Li(2015)}]{dis}
\bibinfo{author}{T.~Li}, \bibinfo{title}{Efficient Virtualization and
  Scheduling for Productive GPU-based High Performance Computing Systems},
  Ph.D. thesis, The George Washington University, \bibinfo{year}{2015}.
\bibitem[{Huang and Hsiung(2009)}]{huang2009hardware}
\bibinfo{author}{C.~Huang}, \bibinfo{author}{P.~Hsiung},
\newblock \bibinfo{title}{Hardware resource virtualization for dynamically
  partially reconfigurable systems},
\newblock \bibinfo{journal}{Embedded Systems Letters, IEEE} \bibinfo{volume}{1}
  (\bibinfo{year}{2009}) \bibinfo{pages}{19--23}.
\bibitem[{Li et~al.(2009)Li, Huang, El-Ghazawi, and Huang}]{rad}
\bibinfo{author}{T.~Li}, \bibinfo{author}{M.~Huang},
  \bibinfo{author}{T.~El-Ghazawi}, \bibinfo{author}{H.~Huang},
\newblock \bibinfo{title}{Reconfigurable active drive: An fpga accelerated
  storage architecture for data-intensive applications},
\newblock \bibinfo{journal}{2009 Symposium on Application Accelerators in
  High-Performance Computing}  (\bibinfo{year}{2009}) \bibinfo{pages}{1--3}.
\bibitem[{El-Araby et~al.(2008)El-Araby, Gonzalez, and
  El-Ghazawi}]{el2008virtualizing}
\bibinfo{author}{E.~El-Araby}, \bibinfo{author}{I.~Gonzalez},
  \bibinfo{author}{T.~El-Ghazawi},
\newblock \bibinfo{title}{{Virtualizing and sharing reconfigurable resources in
  High-Performance Reconfigurable Computing systems}},
\newblock in: \bibinfo{booktitle}{High-Performance Reconfigurable Computing
  Technology and Applications, 2008. HPRCTA 2008. Second International Workshop
  on}, \bibinfo{organization}{IEEE}, \bibinfo{year}{2008}, pp.
  \bibinfo{pages}{1--8}.
\bibitem[{{Khronos OpenCL Working Group}(2013)}]{m_opencl}
\bibinfo{author}{{Khronos OpenCL Working Group}}, \bibinfo{title}{The OpenCL
  Specification V2.0}, \bibinfo{year}{2013}.
\bibitem[{Malik et~al.(2011)Malik, Li, Sharif, Shahid, El-Ghazawi, and
  Newby}]{CPE:CPE1860}
\bibinfo{author}{M.~Malik}, \bibinfo{author}{T.~Li},
  \bibinfo{author}{U.~Sharif}, \bibinfo{author}{R.~Shahid},
  \bibinfo{author}{T.~El-Ghazawi}, \bibinfo{author}{G.~Newby},
\newblock \bibinfo{title}{{Productivity of GPUs under different programming
  paradigms}},
\newblock \bibinfo{journal}{Concurrency and Computation: Practice and
  Experience}  (\bibinfo{year}{2011}) \bibinfo{pages}{179--191}. \URLprefix
  \url{http://dx.doi.org/10.1002/cpe.1860}. \DOIprefix\doi{10.1002/cpe.1860}.
\bibitem[{Bailey et~al.(1991)Bailey, Barszcz, Barton, Browning, Carter, Dagum,
  Fatoohi, Frederickson, Lasinski, Schreiber et~al.}]{bailey1991parallel}
\bibinfo{author}{D.~Bailey}, \bibinfo{author}{E.~Barszcz},
  \bibinfo{author}{J.~Barton}, \bibinfo{author}{D.~Browning},
  \bibinfo{author}{R.~Carter}, \bibinfo{author}{L.~Dagum},
  \bibinfo{author}{R.~Fatoohi}, \bibinfo{author}{P.~Frederickson},
  \bibinfo{author}{T.~Lasinski}, \bibinfo{author}{R.~Schreiber}, et~al.,
\newblock \bibinfo{title}{The {NAS} parallel benchmarks},
\newblock \bibinfo{journal}{International Journal of High Performance Computing
  Applications} \bibinfo{volume}{5} (\bibinfo{year}{1991}) \bibinfo{pages}{63}.
\bibitem[{Black and Scholes(1973)}]{black1973pricing}
\bibinfo{author}{F.~Black}, \bibinfo{author}{M.~Scholes},
\newblock \bibinfo{title}{The pricing of options and corporate liabilities},
\newblock \bibinfo{journal}{The journal of political economy}
  (\bibinfo{year}{1973}) \bibinfo{pages}{637--654}.
\bibitem[{w_v(2013)}]{w_vmd}
\bibinfo{title}{Visual molecular dynamics program webpage},
\newblock \bibinfo{year}{2013}.

\end{thebibliography}
\end{document}